\documentclass[seceq]{ptptex}
\usepackage[dvips]{graphicx}
\usepackage{wrapft}

\def\lesim{\m@thcombine<\sim}
\def\gesim{\m@thcombine>\sim}
\def\lessgtr{\m@thcombine<>}
\def\gtrless{\m@thcombine><}

\newcommand{\bra}[1]{\left\langle #1 \right|}

\newcommand{\ket}[1]{\left| #1 \right\rangle}

\newcommand{\adag}{a^{\dagger}}

\newcommand{\Ahat}{\hat{A}}

\newcommand{\Ahatdag}{\hat{A}^{\dagger}}

\newcommand{\Bhat}{\hat{B}}
\newcommand{\Bhatdag}{\hat{B}^{\dagger}}

\newcommand{\cdag}{c^{\dagger}}

\newcommand{\Hhat}{\hat{H}}

\newcommand{\hhat}{\hat{h}}
\newcommand{\fp}{f^{(+)}}

\newcommand{\fm}{f^{(-)}}
\newcommand{\Fhat}{\hat{F}}
\newcommand{\Fhatdag}{\hat{F}^\dagger}
\newcommand{\Fhatp}{\hat{F}^{(+)}}
\newcommand{\Fhatm}{\hat{F}^{(-)}}

\newcommand{\Hc}{{\cal H}}

\newcommand{\Nhat}{\hat{N}}
\newcommand{\Nt}{\tilde{N}}

\newcommand{\Dhat}{\hat{D}}

\newcommand{\Qhat}{\hat{Q}}
\newcommand{\Rhat}{\hat{R}}
\newcommand{\Phat}{\hat{P}}

\newcommand{\That}{\hat{\Theta}}

\newcommand{\Fp}{F^{(+)}}
\newcommand{\Fm}{F^{(-)}}
\newcommand{\Rp}{R^{(+)}}
\newcommand{\Rm}{R^{(-)}}

\newcommand{\Ab}{\mbox{\boldmath $A$}}
\newcommand{\Abdag}{\mbox{\boldmath $A$}^{\dagger}}

\newcommand{\Nb}{\mbox{\boldmath $N$}}

\newcommand{\del}{\partial}

\newcommand{\beq}{\begin{equation}}
\newcommand{\beqa}{\begin{eqnarray}}
\newcommand{\eeq}{\end{equation}}
\newcommand{\eeqa}{\end{eqnarray}}

\newcommand{\fb}{\mbox {\bfseries\itshape f}}
\newcommand{\SB}{\mbox {\bfseries\itshape S}}

\title{
Gauge-Invariant Formulation of Adiabatic Self-Consistent Collective
Coordinate Method}

\author{Nobuo HINOHARA,$^1$~~ Takashi NAKATSUKASA,$^2$~~ 
Masayuki MATSUO$^3$ \\ 
and~ Kenichi MATSUYANAGI$^1$}
\inst{
{\small\it $^1$ Department of Physics, Graduate School of Science,}\\
{\small\it Kyoto University, Kyoto 606-8502, Japan}\\
{\small\it $^2$ Institute of Physics and Center for Computational Sciences, 
University of Tsukuba,}\\
{\small\it Tsukuba 305-8571, Japan}\\
{\small\it $^3$ Department of Physics, Faculty of Science,}\\
{\small\it Niigata University, Niigata 950-2181, Japan}
}
\abst{
The adiabatic self-consistent collective coordinate (ASCC) method
is a practical microscopic theory
of large-amplitude collective motions in nuclei with superfluidity.
We show that its basic equations are invariant 
against transformations involving the gauge angle in the particle-number space.
By virtue of this invariance, a clean separation between the large-amplitude 
collective motion and the pairing rotational motion can be achieved, 
enabling us to restore the particle-number symmetry broken by 
the Hartree-Fock-Bogoliubov (HFB) approximation.
We formulate the ASCC method explicitly in a gauge-invariant form.
In solving the ASCC equations, it is necessary to fix the gauge.
Applying this new formulation to the multi-$O$(4) model, we compare
different gauge-fixing procedures and demonstrate that calculations 
using different gauges indeed yield the same results for 
gauge-invariant quantities, such as the collective path and quantum spectra. 
We suggest a gauge-fixing prescription that seems most convenient in 
realistic calculations.
}

\begin{document}

\maketitle

\section{Introduction}

Construction of the microscopic theory of large amplitude collective
motion is a long-standing and fundamental subject of nuclear many-body problem.
\cite{row76,bri76,vil77,mar77,bar78,goe78,mar80,gia80,dob81,goe81,muk81,
row82,fio83,rei84,kur83,yam84,mat85,mat86,shi87,yam87,
bul89,wal91,kle91a,kan94,nak98a,nak98b,nak00,lib99,yul99,pro04,alm04a,alm04b,
kle91b,dan00,kur01}.
As is well known, pairing correlations play crucial roles
in low-lying states of medium and heavy mass nuclei, and
they are taken into account in the 
Hartree-Fock-Bogoliubov (HFB) mean-field theory by breaking
the particle number conservation.
\cite{rin80,bla86,abe83}
The broken particle-number symmetry can be restored, however,  by making 
the self-consistent quasiparticle random-phase approximation (QRPA) 
in addition to the HFB mean field.
\cite{rin80,bla86,abe83} 
It is an advantage of the QRPA that number-conserving collective modes,
such as shape-vibrational modes, are
exactly decoupled from number-fluctuation modes. 
The latter modes are associated with the nucleon-number degrees of 
freedom and called the pairing rotational modes.
It is a unique feature of nuclei as finite quantum systems
that such a rotational motion in the gauge space is 
actually observed in quantum spectra.\cite{bri05}
Since the applicability of the QRPA is limited to small-amplitude 
collective motions, it is highly desirable to extend the QRPA to a general
theory keeping its decoupling feature.
Such a theory should be capable of describing 
the interplay between large-amplitude collective motions 
and the pairing rotational modes.

The self-consistent collective coordinate (SCC) method\cite{mar80} is 
a microscopic theory of large-amplitude collective motion 
based on the time-dependent Hartree-Fock (TDHF) method, 
which enables us to extract the collective
submanifold in a fully self-consistent manner.
The SCC method was originally formulated\cite{mar80} for systems without pairing
correlations, and then extended\cite{mat86} to systems with superfluidity.
To extract the collective submanifold embedded in the time-dependent
Hartree-Fock-Bogoliubov (TDHFB) phase space, the number and angle variables 
describing the pairing rotational motion are explicitly introduced\cite{mat86} 
in addition to the collective variables describing the large-amplitude 
collective motion. 
This extended version of the SCC method has been applied successfully to 
various kinds of anharmonic vibration
and high-spin rotational motion.
\cite{mat84,mat85a,mat85b,tak89,yam89a,yam89b,aib90,
yam91,ter91,ter92,mat92,shi01}
However, its solution relied on
an expansion technique with respect to the collective coordinates and momenta
around the HFB states.
Thus, it is difficult to describe collective motions with a genuine
large-amplitude nature.
Recently, the adiabatic SCC (ASCC) method has been proposed to overcome
this difficulty\cite{mat00}.
The ASCC method is an alternative way to solve the SCC basic equations
assuming that the large-amplitude collective motion of interest is 
slow (adiabatic). Under this assumption, the basic equations of the SCC 
method are expanded up to the second order in the collective momentum, 
but no expansion is made with respect to the collective coordinate.
This has been applied\cite{kob04} to shape-coexistence phenomena,
\cite{woo92,fis00,bou03}
where the large-amplitude collective motions 
take place between the oblate and prolate HFB equilibrium shapes.
However, the calculation often leads to a numerical instability,
caused by a redundant degree of freedom in the ASCC equations,
which was treated with an additional condition by hand.\cite{hin06}
Now we understand the origin of this redundancy.
It turns out to be due to a gauge invariance of the ASCC equations.

The main purpose of this paper is to formulate the ASCC method
in a way to manifest the invariance against a transformation
with respect to the angle variable in the gauge space.
This invariance is crucial to achieve a clean 
separation between the large-amplitude collective motions and the pairing 
rotational motions, and to restore the particle-number symmetry broken by the 
HFB approximation. This new formulation of the ASCC method also 
provides a justification to the prescription adopted in our previous work
\cite{hin06}.
We apply the method to the multi-$O(4)$ model
\cite{hin06,mat82,miz81,suz88,fuk91,kob03}
with different choices of the gauge
and test the internal consistency of the proposed scheme by carrying out 
a detailed numerical calculation.

This paper is organized as follows:
In \S\ref{sec:basic-ascc-eq} the basic equations of the ASCC method
are recapitulated.
In \S\ref{sec:gauge} the gauge-invariant formulation of the ASCC method 
is given.
It is applied to the multi-$O(4)$ model in \S\ref{sec:multi-O4-ASCC} 
and tested by numerical calculations in \S\ref{sec:numerical-cal}.
Concluding remarks are given in \S\ref{sec:Conclusions}.

\section{Basic equations of the ASCC method}\label{sec:basic-ascc-eq}

\subsection{Basic ideas}\label{sec:basic-scc}

Time evolution of large amplitude collective motion is described by  
the time-dependent variational principle,
\begin{align}
 \delta \bra{\phi(t)}i\frac{\del}{\del t} - \Hhat \ket{\phi(t)} = 0,
 \label{eq:TDVP}
\end{align} 
for the time-dependent HFB state vector $\ket{\phi(t)}$.
Assuming that the time-dependence of the collective motion is 
governed by the collective coordinate $q$ and the momentum $p$,
we parameterize the time-dependent HFB state vector as
\begin{align}
 \ket{\phi(t)} = \ket{\phi(q,p,\varphi,n)}.
\end{align}
Here, $\varphi$ represents the gauge angle conjugate to
the particle-number $n\equiv N-N_0$. We measure the 
particle number from a reference value $N_0$ specified below,
and assume, for simplicity, that the number of collective coordinate 
is one. 
We define the intrinsic state vector $\ket{\phi(q,p,n)}$ 
in the particle-number (gauge) space by
\begin{align}
 \ket{\phi(q,p,\varphi,n)} = e^{-i\varphi\Nt} \ket{\phi(q,p,n)},
 \label{eq:def_intr}
\end{align}
where $\Nt\equiv \Nhat-N_0$.
Two sets of collective variables $(q,p)$ and $(\varphi,n)$ 
are determined such that the canonical variable conditions,
\begin{align}
 \bra{\phi(q,p,n)}i\frac{\del}{\del q}
 \ket{\phi(q,p,n)} 
 = p + \frac{\del S}{\del q}, \quad &
 \bra{\phi(q,p,n)}\frac{\del}{i\del p}
 \ket{\phi(q,p,n)} = - \frac{\del S}{\del p}, 
 \nonumber \\
 \bra{\phi(q,p,n)}\Nt
 \ket{\phi(q,p,n)} = n+
 \frac{\del S}{\del\varphi}, \quad &
 \bra{\phi(q,p,n)}\frac{\del}{i\del n}
 \ket{\phi(q,p,n)}=-\frac{\del S}{\del n},
 \label{eq:SCC-cvc}
\end{align}
are satisfied. 
Here the generating function $S$ is an arbitrary function of 
$q,p,\varphi$, and $n$. We choose $S=0$ because it is appropriate
to the adiabatic approximation.\cite{kur83,yam84,yam87}
The collective Hamiltonian is defined by
\begin{align}
\Hc (q,p,n) = \bra{\phi(q,p,\varphi,n)}\Hhat\ket{\phi(q,p,\varphi,n)}
= \bra{\phi(q,p,n)}\Hhat\ket{\phi(q,p,n)}.
 \label{eq:SCC-Hcoll}
\end{align}
Note that it is independent of the gauge angle $\varphi$, 
because the Hamiltonian commutes with the particle-number operator $\Nhat$.

The equation of collective path is obtained by replacing the 
time derivative term in Eq.(\ref{eq:TDVP}) with derivatives with respect to 
four collective variables
\begin{align}
 \delta\bra{\phi(q,p,n)}\Hhat - i\left(
\frac{\del\Hc}{\del p}\frac{\del}{\del q} - \frac{\del \Hc}{\del
 q}\frac{\del}{\del p} + \frac{1}{i}\frac{\del \Hc}{\del n}\Nt
\right)
\ket{\phi(q,p,n)} = 0, \label{eq:SCC-collectivepath}
\end{align}
where the canonical equations of motion for collective
variables 
$(q,p)$ and $(\varphi,n)$ are used in order to 
eliminate the time-derivative of the collective variables.
Equations (\ref{eq:SCC-cvc}), (\ref{eq:SCC-Hcoll}), and 
(\ref{eq:SCC-collectivepath}) constitute the basic 
equations of the SCC method.\cite{mar80,mat86}

\subsection{Basic equations of the ASCC method}

Assuming that the large-amplitude collective motion is slow,
i.e., $p$ is small, let us write the TDHFB state vector $\ket{\phi(q,p,n)}$ 
in the following form:
\begin{align}
 \ket{\phi(q,p,n)} 
 =  e^{ip\Qhat(q)+in\That(q)}\ket{\phi(q)}. \label{eq:TDHFBvector}
\end{align}
Here $\Qhat(q)$ and $\That(q)$ are Hermitian one-body operators, 
which may be written as 
\begin{align}
 \Qhat(q) &= \sum_{\alpha\beta} \left( Q_{\alpha\beta}(q) \adag_\alpha
 \adag_\beta + Q_{\alpha\beta}(q)^\ast a_\beta a_\alpha \right), \\
 \That(q) &= i \sum_{\alpha\beta} \left( \Theta_{\alpha\beta}(q) \adag_\alpha
 \adag_\beta - \Theta_{\alpha\beta}(q)^\ast a_\beta a_\alpha\right), 
\end{align}
where $\adag_{\alpha}$ and $a_{\alpha}$ are quasiparticle creation 
and annihilation operators associated with the time-even state vector 
$\ket{\phi(q)}$ and satisfy $a_{\alpha}\ket{\phi(q)} = 0$, 
and $n=N-N_0$, $N_0$ being the expectation value of the particle number
with respect to $\ket{\phi(q)}$.
We shall discuss in \S\ref{sec:gauge} that it is also possible to adopt 
a slightly different representation for $\Qhat(q)$.

Substituting (\ref{eq:TDHFBvector}) into (\ref{eq:SCC-cvc}) 
and comparing the coefficients of the zero-th and the first order terms 
in $p$ and $n$, we obtain the canonical variable conditions 
in the adiabatic limit:
\begin{align}
 \bra{\phi(q)}\Phat(q)\ket{\phi(q)} &= 0, \label{eq:can-p} \\
 \bra{\phi(q)}\Qhat(q)\ket{\phi(q)} &= 0,\label{eq:can-q} \\
 \bra{\phi(q)}\Nt\ket{\phi(q)}    &= 0,\label{eq:can-n} \\
 \bra{\phi(q)}\That(q)\ket{\phi(q)} &= 0,\label{eq:can-t} \\
 \bra{\phi(q)}[\That(q),\Nt]\ket{\phi(q)} &= i, \label{eq:can-nt}\\
 \bra{\phi(q)}[\Qhat(q),\That(q)]\ket{\phi(q)} &= 0, \label{eq:can-qt}\\ 
 \bra{\phi(q)}\frac{\del \Qhat}{\del q}\ket{\phi(q)} &= -1, 
 \label{eq:can-qdel}
\end{align}
where $\Phat(q)$ is the local shift operator defined by
\begin{align}
 \Phat(q)\ket{\phi(q)} = i\frac{\del}{\del q}\ket{\phi(q)}.
\end{align}
Differentiating (\ref{eq:can-q}) and (\ref{eq:can-n}) with respect to 
$q$ and using (\ref{eq:can-qdel}), we obtain 
\begin{align}
 \bra{\phi(q)}[\Qhat(q),\Phat(q)]\ket{\phi(q)} &= i, \label{eq:can-qp} \\
 \bra{\phi(q)}[\Nt, \Phat(q)]\ket{\phi(q)} &= 0.  \label{eq:can-np}
\end{align}

The collective Hamiltonian (\ref{eq:SCC-Hcoll}) is also expanded 
up to the second order in $p$ and the first order in $n$, 
\begin{align}
 \Hc(q,p,n) =& V(q) + \frac{1}{2}B(q)p^2 + \lambda(q)n,
 \label{eq:ASCC-Hcoll}
\end{align}
where the collective potential $V(q)$, the inverse mass parameter $B(q)$, 
and the chemical potential $\lambda(q)$ are defined by
\begin{align}
 V(q) =&  \Hc(q,p,n)\Big\arrowvert_{p=n=0} =
\bra{\phi(q)}\Hhat\ket{\phi(q)}, \\
 B(q) =& \frac{1}{2}\frac{\del^2 \Hc}{\del p^2}
 \Big\arrowvert_{p=n=0} =
 \bra{\phi(q)}[[\Hhat, i\Qhat(q)], i\Qhat(q)]\ket{\phi(q)}, \label{eq:defB}\\
 \lambda(q) =& \frac{\del \Hc}{\del n}\Big\arrowvert_{p=n=0} =
 \bra{\phi(q)}[\Hhat, i\That(q)]\ket{\phi(q)}.
\end{align}

We obtain the ASCC equations by expanding 
the equation of collective path (\ref{eq:SCC-collectivepath}) 
with respect to $p$ and $n$, 
and requiring that the variations vanish for each order in $p$ and $n$. 
In the zero-th order, we obtain the moving-frame HFB equation:
\begin{align}
 \delta\bra{\phi(q)}\Hhat_M(q)\ket{\phi(q)} = 0, \label{eq:ascc1}
\end{align}
where 
\begin{align}
 \Hhat_M(q) = \Hhat - \lambda(q)\Nt - \frac{\del V}{\del q}\Qhat(q),
 \label{eq:ascc1b}
\end{align}
is the moving-frame Hamiltonian.
In the first and the second orders, we obtain
the local harmonic equations (also called the moving-frame QRPA equation):
\begin{align}
\delta\bra{\phi(q)}[\Hhat_M(q), i\Qhat(q) ] - B(q)\Phat(q)
\ket{\phi(q)} = 0, \label{eq:ascc2}
\end{align}
\begin{multline}
\delta\bra{\phi(q)} [\Hhat_M(q), \Phat(q)] - iC(q)\Qhat(q) \\
- \frac{1}{2B(q)}[[\Hhat_M(q), \frac{\del V}{\del q}\Qhat(q)], i\Qhat(q)]
- i\frac{\del\lambda}{\del q} \Nt \ket{\phi(q)} = 0,
\label{eq:ascc3}
\end{multline}
where 
\begin{align}
C(q) = \frac{\del^2 V}{\del q^2}
+ \frac{1}{2B(q)}\frac{\del B}{\del q}\frac{\del V}{\del q}.
\end{align}
Note that in Ref.~\citen{mat00}
the curvature term $1/2B(q)[[\Hhat_M(q),\del V/\del q \Qhat(q)],i\Qhat(q)]$
is linearlized with respect to $\Qhat(q)$ using the relation
\begin{align}
 (\Hhat - \lambda(q)\Nhat)^A = \frac{\del V}{\del q}\Qhat(q), \label{eq:dVdqQ}
\end{align}
where the superscript $A$ in Eq.~(\ref{eq:dVdqQ}) 
denotes the two-quasiparticle creation ($\adag\adag$) 
and annihilation ($aa$) part of the operator in the parenthesis.
Hereafter, we call them ``$A$-part'' and the $\adag a$ terms ``$B$-part''.
The collective variables $(q,p)$ and the collective Hamiltonian
$\Hc(q,p,n)$ are determined by solving the ASCC equations, 
(\ref{eq:ascc1}), (\ref{eq:ascc2}), and (\ref{eq:ascc3}),
under the canonical variable conditions.
Note that we can make a scale transformation of the collective coordinate $q$ 
such that $B(q)=1$. We adopt this choice. Then, $C(q)$ represents
the curvature of the collective potential:
\begin{align}
C(q)=\frac{\partial^2 V(q)}{\partial q^2}.
\label{eq:freq}
\end{align}

\section{Gauge invariance of the ASCC equations with respect to 
the pairing rotational degree of freedom}
\label{sec:gauge}

\subsection{Gauge invariance at the HFB equilibrium point}\label{sec:gauge-1}

As mentioned in the preceding section, the first step to solve the ASCC
equations is to find a solution at one of the HFB equilibrium point
denoted by $q=q_0$, which corresponds to the local minimum
of the collective potential $V(q)$ satisfying $\del V/\del q=0$.
The moving-frame HFB equation reduces to the conventional HFB equation 
at the equilibrium point,
\begin{align}
 \delta\bra{\phi(q_0)}\Hhat - \lambda(q_0)\Nt\ket{\phi(q_0)} = 0. 
 \label{eq:HFB}
\end{align}
The local harmonic equations at the equilibrium point are given by
\begin{align}
 \delta\bra{\phi(q_0)}[\Hhat-\lambda(q_0)\Nt,i\Qhat(q_0)] -
 B(q_0)\Phat(q_0)\ket{\phi(q_0)} &= 0, 
 \label{eq:localharmonic-at-eq1}\\
 \delta\bra{\phi(q_0)}[\Hhat-\lambda(q_0)\Nt,\Phat(q_0)]
 - iC(q_0)\Qhat(q_0) - i\frac{\del\lambda}{\del q}\Nt\ket{\phi(q_0)} &= 0.
 \label{eq:localharmonic-at-eq2}
\end{align}
These equations reduce to the QRPA equations if the quantity
$\del\lambda/\del q$ vanishes. In other words, 
the QRPA solution corresponds to the special solution
of the local harmonic equations with $\del\lambda/\del q = 0$.

Let us consider the following transformations:
\begin{subequations}
\begin{align} 
 \Qhat(q_0) &\rightarrow ~\Qhat(q_0) + \alpha \Nhat^A(q_0), 
 \label{eq:gauge1-Q}\\
 \That(q_0) &\rightarrow ~\That(q_0) + \alpha \Phat(q_0) 
 \label{eq:gauge1-T} \\
 \frac{\del\lambda}{\del q}(q_0) &\rightarrow ~\frac{\del\lambda}{\del q}(q_0) 
 - \alpha C(q_0). \label{eq:gauge1-fN}
\end{align}\label{eq:gauge1}
\end{subequations}
Here,  $\alpha$ is an arbitrary number 
and $\Nhat^A$ denotes the $A$-part of the number operator $\Nhat$.
This is a kind of gauge transformation with respect to the pairing rotational 
degree of freedom.
We can easily confirm that the local harmonic equations 
at the HFB equilibrium point, 
(\ref{eq:localharmonic-at-eq1})-(\ref{eq:localharmonic-at-eq2}), 
and the canonical variable conditions, 
(\ref{eq:can-p})-(\ref{eq:can-qt}) and (\ref{eq:can-qp})-(\ref{eq:can-np}),  
are invariant under this transformation.
Due to this invariance, the solution of the local harmonic equations 
is not uniquely determined at the HFB equilibrium point.
If we choose a value of $\alpha$ such that  
$\del\lambda/\del q =0$ holds, the local harmonic equations 
coincide with the conventional QRPA equations.
We can choose other values of $\alpha$, however, if it is more convenient.

\subsection{Gauge invariance at non-equilibrium points}
\label{sec:ASCC non-eq}

At non-equilibrium points, $\del V/\del q$ is non-zero and 
the moving-frame Hamiltonian 
(\ref{eq:ascc1b}) depends on the collective coordinate operator $\Qhat(q)$. 
Still we can generalize the above consideration at the HFB equilibrium point 
to a general off-equilibrium point $q$ on the collective path: 
It is straightforward to confirm that 
all the basic equations of the ASCC method 
[i.e., the collective Hamiltonian $\Hc(q,p,n)$, (\ref{eq:ASCC-Hcoll}), 
the inverse mass parameter $B(q)$, (\ref{eq:defB}), 
the moving-frame HFB equation, (\ref{eq:ascc1}), 
the local harmonic equations, 
(\ref{eq:ascc2})-(\ref{eq:ascc3}), 
and the canonical variable conditions, 
(\ref{eq:can-p})-(\ref{eq:can-qt}) and (\ref{eq:can-qp})-(\ref{eq:can-np})]   
are invariant under the following transformations 
with respect to the pairing rotational degree of freedom:
\begin{subequations}
\label{eq:gauge-new}
\begin{align}
 \Qhat(q) \rightarrow & ~\Qhat(q) + \alpha\Nt, \\
 \That(q) \rightarrow & ~\That(q) + \alpha\Phat(q), \\
 \lambda(q) \rightarrow & ~\lambda(q) - \alpha \frac{\del V}{\del q}(q),
 \\
 \frac{\del\lambda}{\del q}(q) \rightarrow & ~\frac{\del\lambda}{\del
 q}(q) - \alpha C(q),
\end{align}
\end{subequations}
if the collective coordinate operator $\Qhat(q)$ is constructed such that 
it is exactly commutable with the number operator $\Nhat$,
\begin{align}
 [\Qhat(q), \Nhat] = 0. \label{eq:QNzero}
\end{align}

In association with the above transformations of $\Qhat(q)$ and $\That(q)$, 
the original TDHFB state vector
\begin{align}
 \ket{\phi(q,p,\varphi,n)} 
= e^{-i\varphi\Nt}e^{ip\Qhat(q)}e^{in\That(q)}\ket{\phi(q)}, 
\label{eq:TDHFBvecin-alimit}
\end{align}
is transformed as 
\begin{align}
 \ket{\phi(q,p,\varphi,n)} \rightarrow &
 e^{-i\varphi\Nt}e^{ip(\Qhat(q)+\alpha\Nt)}
  e^{in(\That(q)+\alpha\Phat(q))}\ket{\phi(q)} \nonumber \\
 = & e^{-i(\varphi-\alpha p)\Nt} e^{ip\Qhat(q)}
   e^{in\That(q)} \ket{\phi(q-\alpha n)}. 
\label{eq:TDHFBvectrans}
\end{align}
Here, the relation
\begin{align}
 \ket{\phi(q+\delta q)} = e^{-i\delta q \Phat(q)}\ket{\phi(q)}
\end{align}
is used and the operators, $\That(q)$ and $\Phat(q)$, are treated as 
commutable with each other under the adiabatic approximation.
We also note that the expression (\ref{eq:TDHFBvecin-alimit}) is slightly
different from (\ref{eq:TDHFBvector}); 
the difference between $e^{ip\Qhat(q)+in\That(q)}$
and $e^{ip\Qhat(q)}e^{in\That(q)}$, however, gives rise to only higher 
order contributions, which are ignored in the adiabatic approximation 
under consideration.
We see that the gauge angle $\varphi$ changes to $\varphi-\alpha p$ 
due to the transformation (\ref{eq:gauge-new}).
Thus, hereafter, we briefly call the transformations (\ref{eq:gauge-new}) 
``gauge transformations'', independence of the choice of $\alpha$ 
``gauge invariance'', and choice of the value of $\alpha$  ``gauge fixing.''

The commutability, (\ref{eq:QNzero}), implies that $\Qhat(q)$ is a normal 
one-body operator written in terms of the nucleon creation and annihilation 
operators ($\cdag,c$) in the following form:
\begin{align}
 \Qhat(q) = \sum_{ij} Q_{ij}(q) :\cdag_i c_j:
          \ \ \equiv \Qhat^A(q)+\Qhat^B(q). \label{eq:Q-ph}
\end{align}
Here, the symbol $:~:$ denotes the normal product part 
when $\Qhat(q)$ is written in terms of 
the quasiparticle operators ($\adag, a$)
defined at $\ket{\phi(q)}$, and the coefficients satisfy 
the relation $Q_{ij}(q)=Q_{ji}(q)^\ast$, 
because $\Qhat(q)$ is supposed to be Hermitian.
Note that this $\Qhat(q)$ operator contains 
the $B$-part, $\Qhat^B(q)=\sum_{kl} Q_{kl}^B \adag_k a_l$,
as well as the $A$-part,
$\Qhat^A(q)=\sum_{kl} Q_{kl}^A \adag_k \adag_l + \mbox{h.c.}$
Accordingly, the relation (\ref{eq:dVdqQ}) does not hold 
for this $\Qhat(q)$ operator.

In this way, we arrive at a new formulation of the ASCC method
in which the gauge-invariance (\ref{eq:gauge-new}) is manifest. 
The gauge-invariant ASCC method consists of the basic equations 
which are the same as those in the original ASCC method\cite{mat00} 
except for the use of Eq.~(\ref{eq:ascc3}) and 
the $\Qhat(q)$ operator given in the form of (\ref{eq:Q-ph}).

\subsection{Gauge fixing and numerical algorithm}

The fact that the ASCC equations are invariant against the gauge 
transformations (\ref{eq:gauge-new}) indicates the necessity of choosing a 
particular gauge for numerical computation: 
If the gauge is not fixed, an instability with respect to the gauge degree 
of freedom might occur during the course of numerical calculation.
Let us outline the procedure of gauge fixing and numerical algorithm 
for solving the gauge invariant ASCC equations.
We start the calculation by solving the moving-frame QRPA 
at one of the HFB equilibrium points. 
A solution of the moving-frame QRPA at the HFB equilibrium point 
can be obtained, as discussed in \S \ref{sec:gauge-1}, 
by choosing the gauge $\del\lambda/\del q = 0$.
Hereafter we call this gauge ``QRPA gauge'', because under this 
gauge, the moving-frame QRPA equations at the HFB equilibrium
reduce to the conventional QRPA equations.
As we shall see later, the numerical calculation
using the QRPA gauge encounters a difficulty 
at inflection points of the collective potential $V(q)$.
It is possible, however, to choose another gauge 
that is free from this difficulty.
With the use of the multi-$O$(4) model,
we shall explicitly show in \S\ref{sec:numerical-cal}
how this is done.

Since the moving-frame HFB equation at non-equilibrium points
contains $\Qhat(q)$ that should be determined by the local harmonic
equations, we have to resort to an iterative procedure.
We proceed to the direction of the lowest energy solution of the
moving-frame QRPA, and successively find the solutions
in the following manner.
Suppose that we already obtain the solution at $q'=q-\delta q$ 
where $\delta q$ is the numerical mesh size in computation.
The moving-frame HFB equation at $q$ for the $n$-th iteration 
\begin{align}
 &\delta\bra{\phi^{(n)}(q)}\Hhat_M^{(n)}(q)\ket{\phi^{(n)}(q)} = 0, \\
 \Hhat_M^{(n)}(q) &= \Hhat - \lambda^{(n)}(q)\Nhat - \frac{\del V}{\del q}^{(n)}(q)\Qhat^{(n-1)}(q),
\end{align}
is solved under the following two constraints 
\begin{align}
 \bra{\phi^{(n)}(q)}\Nhat\ket{\phi^{(n)}(q)} &= N_0, \\
 \bra{\phi^{(n)}(q)}\Qhat(q-\delta q)\ket{\phi^{(n)}(q)} &= 
 \delta q, \label{eq:q-const}
\end{align}
which are derived from the canonical variable conditions,
(\ref{eq:can-n}) and (\ref{eq:can-qdel}), respectively.
In starting this iterative procedure at $q$, 
the neighboring solution $\Qhat(q-\delta q)$ (or a linear combination
of the moving-frame QRPA modes at $q-\delta q$) may be used 
as an initial trial for the operator $\Qhat(q)^{(0)}$.
The moving-frame QRPA equations for the $n$-th iteration are
written as
\begin{align}
 \delta\bra{\phi^{(n)}(q)}[\Hhat_M^{(n)}(q), i\Qhat^{(n)}(q)]
 - B^{(n)}(q)\Phat^{(n)}(q) \ket{\phi^{(n)}(q)} = 0,
\end{align}
\begin{multline}
 \delta\bra{\phi^{(n)}(q)}[\Hhat_M^{(n)}(q),
 \Phat^{(n)}(q)]
 - \frac{1}{2B^{(n)}(q)}[[\Hhat_M^{(n)}(q), \frac{\del V}{\del q} \Qhat(q)], i\Qhat(q)] \\
 - iC^{(n)}(q)\Qhat^{(n)}(q) - i\frac{\del\lambda}{\del q}^{(n)}(q)\Nhat
 \ket{\phi^{(n)}(q)} = 0.
\end{multline}
As the curvature term is nonlinear with respect to $\Qhat(q)$,
we replace one of the operator $\Qhat(q)$ with that of the previous
iteration step $\Qhat^{(n-1)}(q)$.
This procedure will be discussed in detail for the multi-$O(4)$ model
in \S\ref{sec:LHE}.
Thus the moving-frame QRPA equations are linearlized with respect
to $\Qhat^{(n)}(q)$ and $\Phat^{(n)}(q)$.
Its $A$-part, $\Qhat^{A(n)}(q)$, can be determined
in a manner similar to solving the conventional QRPA equations.
After obtaining $\Qhat^{A(n)}(q)$,
its $B$-part, $\Qhat^{B(n)}(q)$, is determined with the requirement,
Eq.~(\ref{eq:QNzero}).
This iterative procedure is repeated until we get convergence for 
$\lambda^{(n)}(q), \Qhat^{(n)}(q)$, and $\ket{\phi^{(n)}(q)}$.

\section{Application of the gauge-invariant ASCC  method 
to the multi-$O(4)$ model} 
\label{sec:multi-O4-ASCC}

In the following sections, we apply the scheme formulated above 
to the multi-$O(4)$ model, 
and discuss gauge-fixing conditions suitable
for solving the gauge-invariant ASCC equations.
We discuss excitation spectra and transition strengths in systems with 
definite particle number $N=N_0$, so that we put $n=0$ 
in the following sections.

\subsection{The multi-$O(4)$ model}

The multi-$O(4)$ model Hamiltonian has been used to test the 
validity of microscopic theories of nuclear collective motion
\cite{hin06,mat82,miz81,suz88,fuk91,kob03}.
We employ exactly the same model Hamiltonian as was used in the previous
work \cite{hin06}.
The model is constructed in terms of the generators of the $O(4)$ symmetry,
\begin{align}
\Ahat_j^\dag &= \sum_{m>0} c_{jm}^{\dag} c_{j-m}^{\dag}, &\quad
\Bhat_j^\dag &= \sum_{m>0} {\sigma_{jm}}c_{jm}^{\dag} c_{j-m}^{\dag},\\
\Nhat_j &= \sum_m c_{jm}^{\dag} c_{jm}, &\quad
\Dhat_j &= \sum_m {\sigma_{jm}} c_{jm}^{\dag} c_{jm},
\end{align}
where the nucleon creation and annihilation operators $(\cdag_{jm}, c_{jm})$
are used and the quantity $\sigma_{jm}$ is given by
\begin{align}
{\sigma_{jm}} =
\left \{
\begin{array}{cc}
 1 &  |m| < \Omega_{j}/2, \\
-1 &  |m|  >   \Omega_{j}/2.
\end{array}
\right.
\end{align}
These four operators represent the monopole pair, the (simplified) 
quadrupole pair, the particle number, 
and the (simplified) quadrupole operators for each $j$-shell, respectively.
The model Hamiltonian is written in the following form:
\begin{align}
\Hhat &= \hhat_0 - \frac{1}{2} G_0 (\Ahatdag \Ahat + \Ahat\Ahatdag)
                      - \frac{1}{2} G_2 (\Bhatdag \Bhat + \Bhat\Bhatdag)
                      - \frac{1}{2} \chi {\Dhat}^{2}, \label{eq:o4Hamiltonian}
\\
\hhat_0 &= \sum_j e^0_j \Nhat_j. \nonumber
\end{align}
where 
defined by
\begin{align}
\Ahatdag = \sum_j \Ahatdag_j,~~
\Bhatdag = \sum_j d_j \Bhatdag_j,~~
\Nhat = \sum_j \Nhat_j,~~
\Dhat = \sum_j d_j \Dhat_j,
\end{align}
and $d_{j}$ represents the quadrupole matrix element.
The first term on the right-hand side of Eq.~(\ref{eq:o4Hamiltonian}) is
the single-particle Hamiltonian, giving a spherical single-particle energy
$e^0_j$ for each $j$-shell which possesses
$(2\Omega_j)$-fold degeneracy ($2\Omega_j=2j+1$).
The other terms represent the residual two-body interactions:
the monopole-pairing interaction, the quadrupole-type pairing interaction, 
and the quadrupole-type particle-hole interaction. 
Their interaction strengths are denoted by $G_0$, $G_2$, and $\chi$, 
respectively.
Here, the operators $\Ahat$ and $\Nhat$ are 
the monopole-pair
and the number operators, while $\Bhat$ and $\Dhat$
represent the simplified quadrupole-pair and quadrupole 
particle-hole operators, respectively.

The residual interactions of this model are separable
and we can always write such a separable Hamiltonian in the following form:
\begin{align}
\Hhat = \hhat_0
  - \frac{1}{2}\sum_{s}\kappa_{s}\Fhatp_{s}\Fhatp_{s}
  + \frac{1}{2}\sum_{s}\kappa_{s}\Fhatm_{s}\Fhatm_{s},
\label{eq:separable}
\end{align}
where
\begin{align}
\Fhat_s^{(\pm)} \equiv (\Fhat_s \pm \Fhatdag_s)/2 = \pm
 \Fhat_s^{(\pm)\dagger}.
\end{align}
The superscripts $(\pm)$ indicate the Hermitian or anti-Hermitian nature of
the bilinear operator $\Fhat$.
The multi-$O(4)$ model Hamiltonian under consideration contains three
kinds of residual interactions.
The indices $s=$1, 2, and 3 on the operators
$\Fhat$ and the interaction strengths $\kappa_{s}$
indicate 
the monopole-pairing, the quadrupole-pairing
and the quadrupole particle-hole interactions, respectively:
$\Fhat_{s=1}=A, \Fhat_{s=2}=B$, $\Fhat_{s=3}=\Dhat$,
$\kappa_1=2G_0, \kappa_2=2G_2$, and $\kappa_3=\chi$.

\subsection{Quasiparticle representation}

To solve the ASCC equations, 
it is convenient to use the quasiparticle basis
locally defined with respect to the state $\ket{\phi(q)}$ 
on the collective path. 
For the multi-$O(4)$ model,
the Bogoliubov transformation to the quasiparticle creation and annihilation 
operators, $a_{i}^{\dag}(q)$ and $a_{i}(q)$, satisfying the vacuum condition, 
$a_{i}(q)\ket{\phi(q)}=0$, is written
\begin{align}           
\left(
\begin{array}{c}
      {\adag_{i}}(q)\\
      {a_{-i}}(q)
\end{array}
\right)
 \equiv
\begin{pmatrix}
           u_{i}(q) & -v_{i}(q)  \\
           v_{i}(q) & u_{i}(q) 
\end{pmatrix}
\left(
\begin{array}{c}
      {\cdag_{i}}\\
      {c_{-i}}
\end{array}
\right).
\label{eq:bogoliubov}
\end{align}
Here, the indices $\pm i$ represent the set of angular momentum quantum
numbers $(j,\pm m)$. The Bogoliubov transformation is locally determined 
on the collective path by the moving-frame HFB equation 
for a given collective coordinate operator $\Qhat(q)$. 

Using the quasiparticle bilinear operators
\begin{align}
\Abdag_i(q) &= \adag_i(q)\adag_{-i}(q), \\
\Nb_i(q) &= \adag_i(q)a_i(q) + \adag_{-i}(q)a_{-i}(q),
\end{align}
the nucleon bilinear operators $\Ahatdag_{i}$ and $\Nhat_{i}$ 
are rewritten as 
\begin{align}
\Ahatdag_{i} &= u_{i}(q) v_{i}(q) + u_{i}^2(q) \Abdag_i(q) 
 - v_{i}^{2}(q) {\Ab}_{i}(q)
 - u_{i}(q) v_{i}(q) \Nb_{i}(q), \\
\Nhat_{i} &= 2v_{i}^{2}(q)+2u_{i}(q)v_{i}(q)(\Abdag_{i}(q)+\Ab_{i}(q))
 +(u_{i}^{2}(q)-v_{i}^{2}(q))\Nb_{i}(q).
\end{align}
The quasiparticle bilinear operators $\Abdag_i(q),\Ab_i(q)$,
and $\Nb_i(q)$ satisfy the following commutation relations:
\begin{align}
\left[ \Ab_i(q),\Abdag_{i'}(q) \right]
     &= \delta_{ii'}(1-\Nb_{i}(q)), \\
\left[ \Nb_{i}(q),\Abdag_{i'}(q)\right]
     &= 2\delta_{ii'} \Abdag_{i'}(q).
\end{align}
The particle number $N_0$, the quadrupole deformation $D(q)$, 
the monopole-pairing gap $\Delta_0(q)$, 
and the quadrupole-pairing gap $\Delta_2(q)$ are given by 
the expectation values with respect to the 
mean-field state vector $\ket{\phi(q)}$:
\begin{align}
N_0  & =  \bra{\phi(q)}\Nhat\ket{\phi(q)} = 2\sum_{i>0} v_{i}^{2}(q),
\label{eq:number}\\
D(q) & =  \bra{\phi(q)}\Dhat\ket{\phi(q)}
       =  2\sum_{i>0} d_{i} \sigma_{i}v_{i}^{2}(q), 
\label{eq:deformation} \\
\Delta_0(q) &= G_0 \bra{\phi(q)}\Ahatdag\ket{\phi(q)} = G_0 \sum_{i>0}
 u_{i}(q)v_{i}(q),
\label{eq:delta0} \\
\Delta_2(q) &= G_2 \bra{\phi(q)}\Bhatdag\ket{\phi(q)} = G_2 \sum_{i>0}
d_{i} \sigma_{i} u_{i}(q) v_{i}(q).
\label{eq:delta2}
\end{align}

Below, we often omit the $q$-dependence in expressions, 
for example, writing $\Ab_i(q)$ as $\Ab_i$.
It should be kept in mind, however, that these quantities are
locally defined with respect to the quasiparticle vacuum $\ket{\phi(q)}$
and depend on $q$.

\subsection{The ASCC equations for separable interactions}

The ASCC equations for the separable Hamiltonian are given by
\begin{align}
\delta\bra{\phi(q)}\hhat_M(q) \ket{\phi(q)} = 0,
\label{eq:ascc1_sep}
\end{align}
\begin{align}
\delta\bra{\phi(q)}[\hhat_M(q), \Qhat(q) ] - \sum_s f^{(-)}_{Q,s} \Fhatm_s 
- \frac{1}{i} B(q) \Phat(q) 
     \ket{\phi(q)} = 0,
\label{eq:ascc2_sep}
\end{align}
\begin{align}
\delta\bra{\phi(q)} \left[\hhat_M(q), \frac{1}{i}B(q)\Phat(q)\right] 
    &- \sum_s f^{(+)}_{P,s}(q) \Fhatp_s 
    - B(q)C(q)\Qhat(q)
    - \sum_s f^{(+)}_{R,s}(q) \Fhatp_s \nonumber \\
&-\frac{1}{2}[[\hhat_M(q), \frac{\del V}{\del q}\Qhat(q)], \Qhat(q)]
 \nonumber \\
    &+ \sum_s \left[\Fhatm_s, \frac{\del V}{\del q}\Qhat(q)\right] f^{(-)}_{Q,s} 
    -f_N(q) \Nt
    \ket{\phi(q)} =0, 
\label{eq:ascc3_sep}
\end{align}
where $\hhat_M(q)$ denotes the self-consistent mean-field Hamiltonian 
in the moving frame, defined by
\begin{equation}
\hhat_M(q) = \hhat(q)  - \lambda(q)\Nt
                       - \frac{\del V}{\del q}\Qhat(q),
\end{equation}
with
\begin{align}
\hhat(q) &= \hhat_0  
- \sum_s \kappa_s \Fhatp_s \bra{\phi(q)}\Fhatp_s\ket{\phi(q)} .
\end{align}
We also define the following quantities
\begin{align}
 &\fm_{Q,s}(q) = -\kappa_s 
              \bra{\phi(q)}[\Fhatm_s, \Qhat(q)] \ket{\phi(q)}, \label{eq:fQs}\\
 &\fp_{P,s}(q)= 
   \kappa_s \bra{\phi(q)}[\Fhatp_s,\frac{1}{i} B(q)\Phat(q)] \ket{\phi(q)}, 
   \label{eq:fPs}\\
 &\fp_{R,s}(q)= -\frac{1}{2}\kappa_s 
 \bra{\phi(q)}\left[\left[\Fhatp_s,\frac{\del V}{\del q}\Qhat(q)\right],\Qhat(q)\right] 
 \ket{\phi(q)}, \label{eq:fRs}\\
 &f_N(q) = B(q)\frac{\del \lambda}{\del q}. \label{eq:fN}
\end{align}
Note that all matrix elements are real, 
so that $\bra{\phi(q)}\Fhatm_s\ket{\phi(q)}=0$.
The fifth term in Eq.(\ref{eq:ascc3_sep}) appears only in the 
gauge-invariant formulation of the ASCC equations because 
there is contribution from this term only if the $\Qhat(q)$ 
contains the $B$-part.

\subsection{The moving-frame HFB equation}

The moving-frame HFB equation (\ref{eq:ascc1_sep}) at a given $q$
determines the time-even TDHB state vector $\ket{\phi(q)}$. 
The variation in Eq.~(\ref{eq:ascc1_sep}) is taken with respect to
arbitrary two quasiparticle states:
\begin{align}
 \delta\ket{\phi(q)} = \adag_{i}(q)\adag_{j}(q)\ket{\phi(q)}.
\end{align}
If we know the operator $\Qhat(q)$,
we can solve this equation using the gradient method so as to 
eliminate the two-quasiparticle terms proportional to $\Abdag_i$ and $\Ab_i$.
The quantities, $\lambda(q)$ and $\del V/\del q$, 
can be regarded as Lagrange multipliers which are
determined by the following two constraints.
The first is the particle number constraint given by (\ref{eq:number}).
This constraint specifies the location in the particle number space.
The second constraint is written as (\ref{eq:q-const}).
For the $\Qhat$ operator defined by (\ref{eq:Q-ph}),
this equation yields
\begin{align}
 \bra{\phi(q)}\Qhat(q-\delta q)\ket{\phi(q)} = 
 2 \sum_{i>0} Q_i(q-\delta q) (v_i(q)^2 - v_i(q-\delta q)^2)= \delta q
\end{align}

\subsection{The local harmonic equations}
\label{sec:LHE}

We solve the local harmonic equations to obtain operators $\Qhat(q)$
and $\Phat(q)$. The collective coordinate operator $\Qhat(q)$ 
is written as 
\begin{align}
 \Qhat(q) = \sum_i Q_i(q) :\Nhat_i: = \sum_{i>0}\left\{ Q_i^A(q)(\Abdag_i 
 + \Ab_i) + Q_i^B(q)\Nb_i \right\}, \label{eq:Qinqp}
\end{align}
while the collective momentum operator $\Phat(q)$ is expressed as 
\begin{align}
 \Phat(q) = i\sum_{i>0} P_i(q) (\Abdag_i - \Ab_i). \label{eq:Pinqp}
\end{align}
As mentioned in the preceding section, 
the $B$-part of the operator $\Phat(q)$ is unnecessary 
in the second order with respect to the collective momentum $p$.

We solve the local harmonic equations in the following way:
Assume that the solution $\Qhat^{(n-1)}(q)$, 
obtained in the previous iteration step, of the local harmonic equations
and the solutions, such as $\ket{\phi(q)}$ and $V(q)$, 
of the moving-frame HFB equation are available. 
(The superscript $n$ is omitted except for $\Qhat(q)$.)
In solving the local harmonic equations, we note that the moving-frame 
Hamiltonian $\hhat_{M}(q)$ and the operators $\Fhat^{(\pm)}_s$ are 
expressed in terms of the quasiparticle bilinear operators 
$\Abdag_i, \Ab_i$, and $\Nb_i$ as
\begin{align}
\hhat_{M}(q) &= V(q) + \sum_{i>0}E_{i}(q){\Nb}_{i}, \label{eq:hm} \\
\Fhatp_{s} &= \bra{\phi(q)}\Fhatp_{s}\ket{\phi(q)} + 
 \Fhatp_{A,s} + \Fhatp_{B,s} \nonumber \\
&= \bra{\phi(q)} \Fhatp_{s} \ket{\phi(q)}
+\sum_{i>0} \Fp_{A,s}(i)
(\Abdag_{i} + \Ab_{i})+\sum_{i>0} \Fp_{B,s}(i){\Nb}_{i},
 \label{eq:Fhatp} \\
\Fhatm_s  &= \sum_{i>0} \Fm_{A,s}(i)(\Abdag_{i}-\Ab_{i})\label{eq:Fhatm}.
\end{align}
Here, 
\begin{align}  
 \Fp_{A,1}(i)&=\frac{1}{2}(u_{i}^{2}-v_{i}^{2}), &
 \Fp_{A,2}(i)&=\frac{1}{2}d_{i} \sigma_{i} (u_{i}^{2} - v_{i}^{2}), &
 \Fp_{A,3}(i)&=2d_{i}\sigma_{i}u_{i}v_{i}, \\  
 \Fm_{A,1}(i)&=-\frac{1}{2}, &
 \Fm_{A,2}(i)&=-\frac{1}{2}d_{i}\sigma_{i}, &
 \Fm_{A,3}(i)&=0, \\ 
 \Fp_{B,1}(i)&=-u_{i}v_{i},&
 \Fp_{B,2}(i)&=-d_{i}\sigma_{i}u_{i}v_{i}, &
 \Fp_{B,3}(i)&= d_{i}\sigma_{i}(u_{i}^{2}-v_{i}^{2}),
\end{align}
\begin{align}
 E_i(q) = (u^2_i - v^2_i)\left(e_i - \chi d_i\sigma_i
 D(q)-\lambda(q)- \frac{\del V}{\del q}Q^{(n-1)}_i(q)\right)
-2(\Delta_0(q)+d_i\sigma_i\Delta_2(q))u_i v_i.
\end{align}
These quantities are determined 
by solving the moving-frame HFB equation (\ref{eq:ascc1_sep}).
For later convenience, we define the following quasiparticle bilinear operators:
\begin{align}
 \Rhat_s^{(\pm)}  \equiv 
 [ \Fhat_{s}^{(\pm)}, \frac{\del V}{\del q}\Qhat^{(n-1)}(q)] 
                = 2 \sum_{i>0}R_{A,s}^{(\pm)}(i) ({\Abdag_i} \mp
 {\Ab_i}) ,\label{eq:Rs}
\end{align}
with
\begin{align}
R_{A,s}^{(+)}(i) &= \frac{\del V}{\del q}\left(
 F_{B,s}^{(+)}(i)Q^{A(n-1)}_i(q) - F_{A,s}^{(+)}(i)Q_i^{B(n-1)}(q)
\right). \\
R_{A,s}^{(-)}(i) &= -\frac{\del V}{\del q} F_{A,s}^{(-)}(i) Q_i^{B(n-1)}(q)
\end{align}
We can express 
the matrix elements $Q^{A(n)}_i$ and $P_i$ in terms 
of $\fm_{Q,s}$, $\fp_{P,s}$, $\fp_{R,s}$ and $f_N$ 
by substituting Eqs.~(\ref{eq:Qinqp}) and (\ref{eq:Pinqp}) 
into Eqs.~(\ref{eq:ascc2_sep}) and (\ref{eq:ascc3_sep}):
\begin{align}
Q^{A(n)}_i =& \frac{2E_{i}}{(2E_{i})^{2}-\omega^2+2\frac{\del V}{\del q}E_iQ_i^{B(n-1)}}
\sum_{s}
 F_{A,s}^{(-)}(i)\fm_{Q,s} \nonumber \\ 
 &+\frac{1}{(2E_{i})^{2}-\omega^2+2\frac{\del V}{\del q}E_iQ_i^{B(n-1)}}
\left\{ \sum_{s}\left(F_{A,s}^{(+)}(i)f_{PR,s}^{(+)} -2 \Rm_{A,s}(i) \fm_{Q,s}\right)
+ N_{i}f_{N}\right\}, 
\label{eq:Qi} \\
P_i =& \frac{2E_{i}}{(2E_{i})^{2}-\omega^2 +2\frac{\del V}{\del q}E_iQ_i^{B(n-1)}}
\left\{ \sum_{s} \left( F_{A,s}^{(+)}(i)f_{PR,s}^{(+)}-2\Rm_{A,s}(i)
 \fm_{Q,s}\right) + N_{i}f_{N}\right\} \nonumber \\ &+ 
\frac{\omega^2-2\frac{\del V}{\del
 q}E_iQ_i^{B(n-1)}}{(2E_{i})^{2}-\omega^2+2\frac{\del V}{\del q}E_iQ_i^{B(n-1)}}\sum_{s} F_{A,s}^{(-)}(i)\fm_{Q,s},
\label{eq:Pi}
\end{align}
where
\begin{align}
&N_i = 2 u_i(q) v_i(q), \\
&f_{PR,s}^{(+)}=f_{P,s}^{(+)}(q)+f_{R,s}^{(+)}(q), \\
&\omega =\sqrt{B(q)C(q)}.
\end{align}
Substituting Eqs.~(\ref{eq:Qinqp}), (\ref{eq:Pinqp}) and
(\ref{eq:Rs}) into Eqs.~(\ref{eq:fQs}), (\ref{eq:fPs}) and (\ref{eq:fRs}),
we obtain
\begin{align}
\fm_{Q,s} &= 2\kappa_{s} \sum_{i>0} \Fm_{A,s}(i)Q^{A(n)}_i,  
\label{eq:disp1} \\
\fp_{PR,s} &= 2\kappa_{s} \sum_{i>0} \left\{ \Fp_{A,s}(i)P_i+
\Rp_{A,s}(i)Q^{A(n)}_i\right\} .
\label{eq:disp2}
\end{align}
Note that $\fm_{Q,3}=0$.
From the canonical variable condition, 
the orthogonality of 
the collective and number fluctuation modes
is required:
\begin{align}
\langle\phi(q)|[\Nt,\Phat(q)]|\phi(q)\rangle 
          = 2i\sum_{i>0} N_i P_i = 0. \label{eq:disp3}
\end{align}
Eliminating $Q^{A(n)}_i$ and $P_i$ from
Eqs.~(\ref{eq:disp1}), (\ref{eq:disp2}), and (\ref{eq:disp3}) 
with use of Eqs.~(\ref{eq:Qi}) and (\ref{eq:Pi}),
we finally obtain the 
dispersion equation
\begin{align}
 \SB (\omega^2)\cdot \fb = 0, \label{eq:disp}
\end{align}
for the quantity
$\fb=\fb(q)=\{\fm_{Q,1}$, $\fm_{Q,2}$, $ \fp_{PR,1}$, $\fp_{PR,2}$, 
$\fp_{PR,3}$, $f_N\}$.
Here $\SB=\{S_{ij}\}$ is a $6\times 6$ matrix whose elements are given by
\begin{subequations}
\begin{align}
 S_{11} &= 2 \kappa_1 \left\{ S^{(1)}(\Fm_{A,1}, \Fm_{A,1}) - 2
 S^{(2)}(\Fm_{A,1}, \Rm_{A,1}) \right\} - 1, \\
 S_{12} &= 2 \kappa_1 \left\{ S^{(1)}(\Fm_{A,1}, \Fm_{A,2}) - 2
 S^{(2)}(\Fm_{A,1}, \Rm_{A,2}) \right\}, \\
 S_{13} &= 2 \kappa_1 S^{(2)}(\Fm_{A,1}, \Fp_{A,1}), \\
 S_{14} &= 2 \kappa_1 S^{(2)}(\Fm_{A,1}, \Fp_{A,2}), \\
 S_{15} &= 2 \kappa_1 S^{(2)}(\Fm_{A,1}, \Fp_{A,3}), \\
 S_{16} &= 2 \kappa_1 S^{(2)}(\Fm_{A,1}, N), 
\end{align}
\end{subequations}
\begin{subequations}
\begin{align}
 S_{21} &= 2 \kappa_2 \left\{ S^{(1)}(\Fm_{A,2}, \Fm_{A,1}) -
 2S^{(2)}(\Fm_{A,2}, \Rm_{A,1}) \right\}, \\
 S_{22} &= 2 \kappa_2 \left\{ S^{(1)}(\Fm_{A,2}, \Fm_{A,2}) -  
 2S^{(2)}(\Fm_{A,2}, \Rm_{A,2}) \right\} - 1,\\
 S_{23} &= 2 \kappa_2 S^{(2)}(\Fm_{A,2}, \Fp_{A,1}), \\
 S_{24} &= 2 \kappa_2 S^{(2)}(\Fm_{A,2}, \Fp_{A,2}), \\
 S_{25} &= 2 \kappa_2 S^{(2)}(\Fm_{A,2}, \Fp_{A,3}), \\
 S_{26} &= 2 \kappa_2 S^{(2)}(\Fm_{A,2}, N), 
\end{align}
\end{subequations}
\begin{subequations}
\begin{align}
 S_{31} &= 2 \kappa_1 \left\{ 
\omega^2 S^{(2)}(\Fp_{A,1},\Fm_{A,1})  - S^{(1)}(\Fp_{A,1}, \Rm_{A,1})
+ S^{(1)}(\Rp_{A,1}, \Fm_{A,1}) - 2 S^{(2)}(\Rp_{A,1}, \Rm_{A,1})
\right\}, \\
 S_{32} &= 2 \kappa_1 \left\{ 
\omega^2 S^{(2)}(\Fp_{A,1},\Fm_{A,2})  - S^{(1)}(\Fp_{A,1}, \Rm_{A,2})
+ S^{(1)}(\Rp_{A,1}, \Fm_{A,2}) - 2 S^{(2)}(\Rp_{A,1}, \Rm_{A,2})
\right\}, \\
 S_{33} &= 2 \kappa_1 \left\{ S^{(1)}(\Fp_{A,1}, \Fp_{A,1}) + S^{(2)}(\Rp_{A,1},
 \Fp_{A,1})\right\} - 1, \\
 S_{34} &= 2 \kappa_1 \left\{ S^{(1)}(\Fp_{A,1}, \Fp_{A,2}) + S^{(2)}(\Rp_{A,1},
 \Fp_{A,2})\right\}, \\
 S_{35} &= 2 \kappa_1 \left\{ S^{(1)}(\Fp_{A,1}, \Fp_{A,3}) + S^{(2)}(\Rp_{A,1},
 \Fp_{A,3})\right\}, \\
 S_{36} &= 2 \kappa_1 \left\{ S^{(1)}(\Fp_{A,1}, N        ) + S^{(2)}(\Rp_{A,1},
 N)\right\}, 
\end{align}
\end{subequations}
\begin{subequations}
\begin{align}
 S_{41} &= 2 \kappa_2 \left\{ \omega^2 S^{(2)}(\Fp_{A,2},\Fm_{A,1}) 
- S^{(1)}(\Fp_{A,2},\Rm_{A,1}) + S^{(1)}(\Rp_{A,2},\Fm_{A,1}) - 2 S^{(2)}(\Rp_{A,2},\Rm_{A,1})
\right\}, \\
 S_{42} &= 2 \kappa_2 \left\{ 
\omega^2 S^{(2)}(\Fp_{A,2},\Fm_{A,2}) - S^{(1)}(\Fp_{A,2}, \Rm_{A,2})
+ S^{(1)}(\Rp_{A,2}, \Fm_{A,2}) - 2 S^{(2)}(\Rp_{A,2}, \Rm_{A,2})
\right\}, \\
 S_{43} &= 2 \kappa_2 \left\{ S^{(1)}(\Fp_{A,2}, \Fp_{A,1}) + S^{(2)}(\Rp_{A,2},
 \Fp_{A,1})\right\}, \\
 S_{44} &= 2 \kappa_2 \left\{ S^{(1)}(\Fp_{A,2}, \Fp_{A,2}) + S^{(2)}(\Rp_{A,2},
 \Fp_{A,2})\right\} - 1, \\
 S_{45} &= 2 \kappa_2 \left\{ S^{(1)}(\Fp_{A,2}, \Fp_{A,3}) + S^{(2)}(\Rp_{A,2},
 \Fp_{A,3})\right\}, \\
 S_{46} &= 2 \kappa_2 \left\{ S^{(1)}(\Fp_{A,2}, N) + S^{(2)}(\Rp_{A,2},
 N)\right\}, 
\end{align}
\end{subequations}
\begin{subequations}
\begin{align}
 S_{51} &= 2 \kappa_3 \left\{ 
\omega^2 S^{(2)}(\Fp_{A,3},\Fm_{A,1}) - S^{(1)}(\Fp_{A,3}, \Rm_{A,1})
+ S^{(1)}(\Rp_{A,3}, \Fm_{A,1}) - 2 S^{(2)}(\Rp_{A,3}, \Rm_{A,1})
\right\}, \\
 S_{52} &= 2 \kappa_3 \left\{ 
\omega^2 S^{(2)}(\Fp_{A,3},\Fm_{A,2}) - S^{(1)}(\Fp_{A,3}, \Rm_{A,2})
+ S^{(1)}(\Rp_{A,3}, \Fm_{A,2}) - 2 S^{(2)}(\Rp_{A,3}, \Rm_{A,2})
\right\}, \\
 S_{53} &= 2 \kappa_3 \left\{ S^{(1)}(\Fp_{A,3}, \Fp_{A,1}) + S^{(2)}(\Rp_{A,3},
 \Fp_{A,1})\right\}, \\
 S_{54} &= 2 \kappa_3 \left\{ S^{(1)}(\Fp_{A,3}, \Fp_{A,2}) + S^{(2)}(\Rp_{A,3},
 \Fp_{A,2})\right\}, \\
 S_{55} &= 2 \kappa_3 \left\{ S^{(1)}(\Fp_{A,3}, \Fp_{A,3}) + S^{(2)}(\Rp_{A,3},
 \Fp_{A,3})\right\} - 1,\\
 S_{56} &= 2 \kappa_3 \left\{ S^{(1)}(\Fp_{A,3}, N) + S^{(2)}(\Rp_{A,3},
 N)\right\}, 
\end{align}
\end{subequations}
\begin{subequations}
\begin{align}
 S_{61} &= \omega^2 S^{(2)}(N, \Fm_{A,1}) - S^{(1)}(N,\Rm_{A,1}), &\quad
 S_{62} &= \omega^2 S^{(2)}(N, \Fm_{A,2}) - S^{(1)}(N,\Rm_{A,2}), \\
 S_{63} &= S^{(1)}(N, \Fp_{A,1}), &\quad
 S_{64} &= S^{(1)}(N, \Fp_{A,2}), \\
 S_{65} &= S^{(1)}(N, \Fp_{A,3}), &\quad
 S_{66} &= S^{(1)}(N, N). 
\end{align}
\end{subequations}
Here, the quantities $S^{(1)}$ and $S^{(2)}$ are defined by
\begin{align}
 S^{(1)}(X,Y) &= \sum_{i>0} \frac{2E_i(q)}{(2E_i(q))^2 - \omega^2(q) +
 2\frac{\del V}{\del q}E_iQ_i^{(B(n-1))}} X_i
 Y_i, \\
 S^{(2)}(X,Y) &= \sum_{i>0} \frac{1}{(2E_i(q))^2 -
 \omega^2(q)+2\frac{\del V}{\del q}E_iQ_i^{(B(n-1))}} X_i Y_i.
\end{align}
The unknown quantities in the dispersion equation (\ref{eq:disp})
are $\fb(q)$ and $\omega^2(q)$.
The squared frequency $\omega^2(q)$ can be determined by the 
condition that the matrix $\SB(\omega^2(q))$ has no inverse:
\begin{align}
\det \SB(\omega^2(q)) = 0. \label{eq:det}
\end{align}
In the case that
there are many solutions $\omega^2(q)$ satisfying this equation,
we choose the smallest of these (including negative values) 
as the collective mode.
Once the value of $\omega^2(q)$ and, consequently, 
the matrix  $\SB(q)$ is specified,
the direction of the vector $\fb(q)$ is found. 
Then,
its absolute value is fixed 
by the normalization condition for the collective mode, i.e.,
\begin{align}
\bra{\phi(q)}[\Qhat^{(n)}(q),\Phat(q)]\ket{\phi(q)} = 2i\sum_{i>0} Q^{A(n)}_i(q)P_i(q) = i.
\end{align}
The choice of the signs of
$\Qhat^{(n)}(q)$ and $\Phat(q)$ are still arbitrary.
This sign specifies the ``rear'' and ``front'' of the one-dimensional
collective path. 

The $B$-part of $\Qhat(q)$ is automatically determined in terms of 
its $A$-part according to its definition (\ref{eq:Qinqp}):
\begin{align}
 Q^{B(n)}_i(q) = \frac{u_i^2-v_i^2}{2u_iv_i}Q_i^{A(n)}(q).
\end{align}

\subsection{Gauge fixing}

Under the gauge transformation (\ref{eq:gauge-new}),
the quantities, $\fm_{Q,1}(q), \fm_{Q,2}(q)$, and $f_N(q)$  
appearing in the dispersion equation (\ref{eq:disp}) transform as
\begin{align}
 \fm_{Q,1}(q) \rightarrow & \fm_{Q,1}(q) - 4\alpha\Delta_0(q),
\label{eq:gauge-f_Q1}\\
 \fm_{Q,2}(q) \rightarrow & \fm_{Q,2}(q) - 4\alpha\Delta_2(q),
\label{eq:gauge-f_Q2}\\ 
 f_N(q)       \rightarrow & f_N(q) - \alpha\omega^2(q). 
\label{eq:gauge-f_N}
\end{align}
These properties clearly indicate that one of the above three quantities 
can be eliminated:
By choosing an appropriate value for $\alpha$ (gauge fixing), 
we can reduce the dimension of the dispersion equation (\ref{eq:disp}) 
to a $5\times 5$ matrix equation.
Namely, Eq.~(\ref{eq:disp}) is redundant and 
the gauge fixing is equivalent to the reduction of its dimension.
The QRPA gauge corresponds to setting $f_N(q)=0$.
In our previous paper~\cite{hin06} we set $\fm_{Q,1}(q)=0$ 
which corresponds to another gauge.
Because the quantity $\fm_{Q,1}(q)$ represents the contribution from the 
time-odd component of the monopole-pairing interaction, 
let us call this gauge ``ETOP (eliminating time-odd pairing) gauge''

\subsection{Requantization}

The solution of the ASCC equations yields the classical collective
Hamiltonian:
\begin{align}
 \Hc(q,p) = \frac{1}{2}p^2 + V(q).
\end{align}
We then obtain the quantum collective Hamiltonian by carrying out the canonical quantization
$\Hc(q,p)\to\Hc\left(q,\displaystyle\frac{1}{i}\frac{\del}{\del q}\right)$.
Note that, in this quantization step, 
there is no ambiguity associated with the ordering of $q$ and $p$, 
because the coordinate scale is chosen 
such that the inverse mass function is unity, i.e., $B(q)=1$.

\section{Numerical test of internal consistency of the proposed scheme}
\label{sec:numerical-cal}

\subsection{Details of numerical calculation}
\label{sec:parameters}
We solve the gauge-invariant ASCC equations for the multi-$O(4)$ model
with the same parameters as in the previous papers~\cite{kob03,hin06}. 
The system consists of 28 particles (one kind of Fermion).
The model space consists of three $j$-shells, labeled $j_1, j_2, j_3$,
with pair degeneracies 
$\Omega_{j_i}=14, \Omega_{j_2}=10, \Omega_{j_3}=4$,
single-particle energies
$e_{j_1}=0, e_{j_2}=1.0, e_{j_3}=3.5$, and
the single-particle quadrupole moments 
$d_{j_1} = 2,\ d_{j_2} = d_{j_3} = 1 $.
Within this model space, the deformation $D=\bra{\phi(q)}\Dhat\ket{\phi(q)}$ 
ranges from $D_{\rm min}=-42$ to $D_{\rm max}=42$.
The calculation is done for the quadrupole-interaction strength $\chi=0.04$ 
and the monopole-pairing-interaction strengths $G_0=0.14, 0.16$ and 0.20.
The properties of the system change from the double-well situation
($G_0=0.14$) to the spherical vibrator ($G_0=0.20$) according to the value of
$G_0$. The effect of the quadrupole pairing is studied 
by comparing the results for $G_2 = 0.00, 0.02$ and 0.04.
As pointed out in Ref.~\citen{hin06} the quadrupole pairing gives significant 
effects on the collective mass.  It is not essential, however,  
for the discussion on the gauge-fixing condition. 
Therefore, we present the results for $G_2 = 0$ in the next subsection 
and show its effect in the final subsection only.
The calculation starts from one of the HFB equilibrium state 
labeled by $q=0$ (see Fig.~\ref{fig:q-D}).
For the deformed cases ($G_0=0.14$ and 0.16),
the HFB equilibrium state having positive (prolate) deformation is 
chosen as a starting point.

\subsection{Comparison of the two gauge fixing conditions}
\label{sec:gauge-compare}

Let us examine whether or not we can find the gauge independent solution 
of the ASCC equations. Existence of the collective path 
that simultaneously satisfies all equations of the ASCC method 
is not self-evident. The aim of the numerical calculation here is to check 
internal consistency of the equations set up in the preceding section.
We solve the gauge-invariant ASCC equations with two different 
gauge fixing conditions:
the QRPA gauge ($f_N(q)=0$) and the ETOP gauge ($\fm_{Q,1}(q)=0$).
In the QRPA gauge, the chemical potential $\lambda(q)$ along the collective
path is set to be constant, while in the ETOP gauge, 
the time-odd contribution of the monopole pairing interaction is fully 
eliminated from the ASCC equations.

\begin{figure}[htbp]
\begin{center}
\begin{tabular}{cc}
\includegraphics[width=65mm]{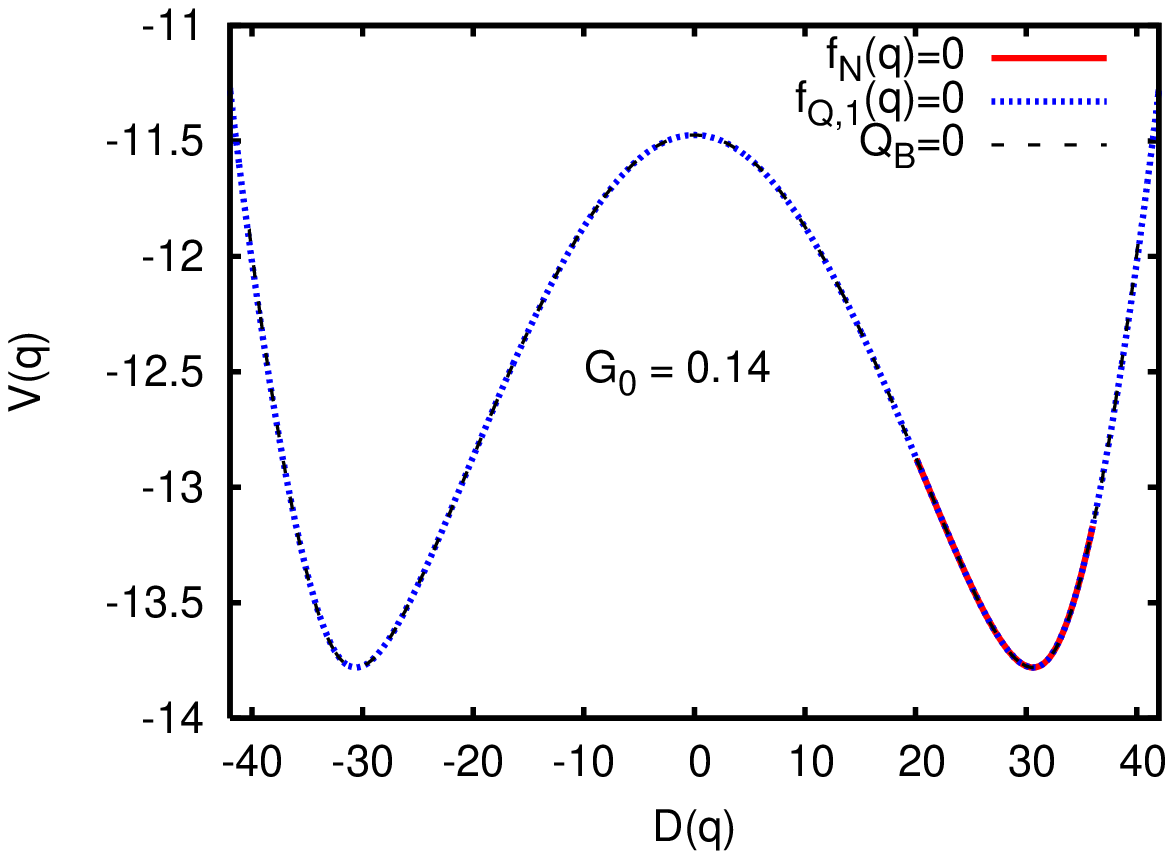} &
\includegraphics[width=65mm]{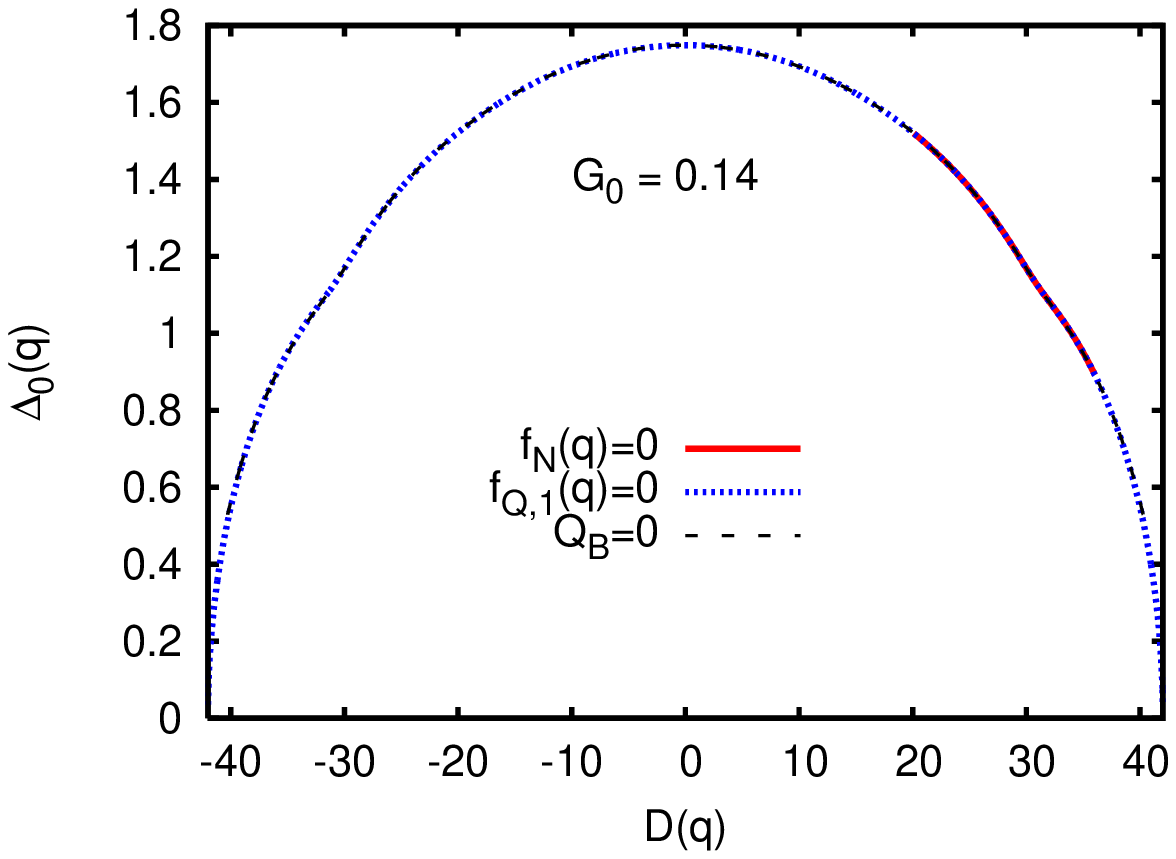} \\
\includegraphics[width=65mm]{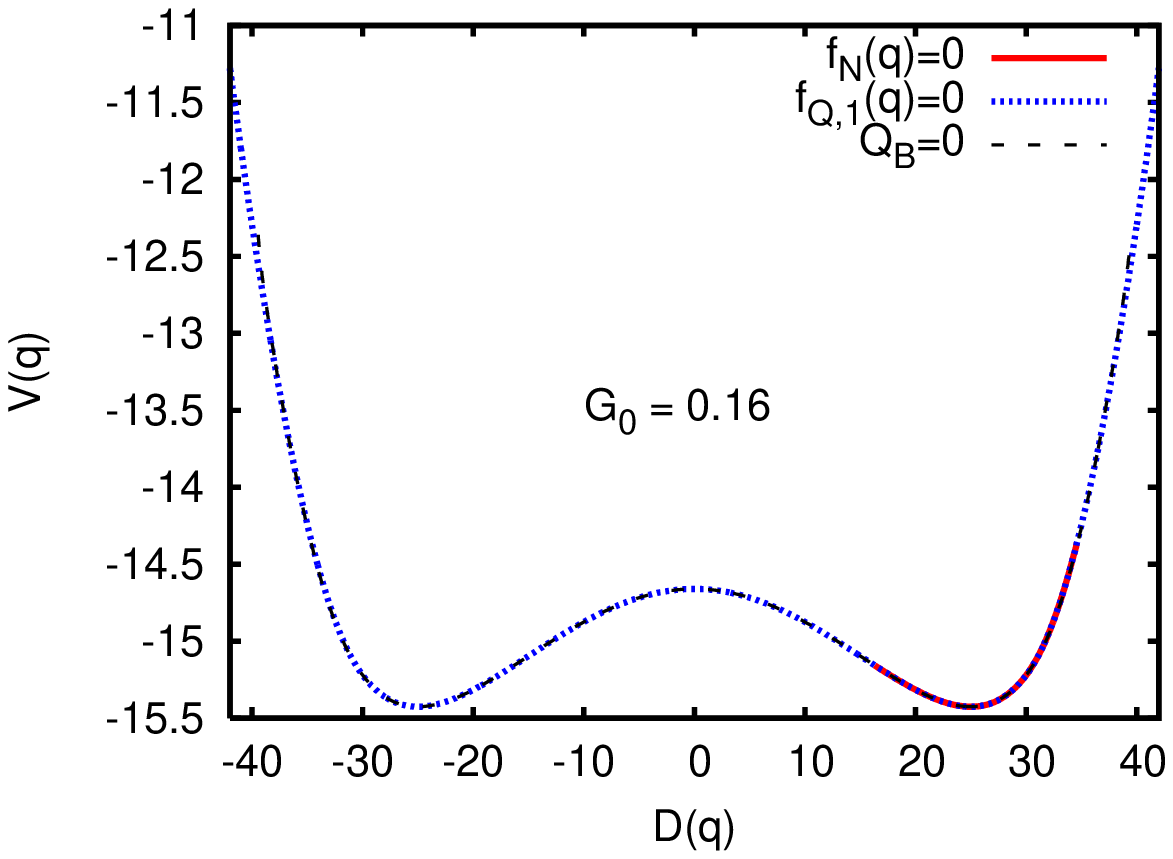} &
\includegraphics[width=65mm]{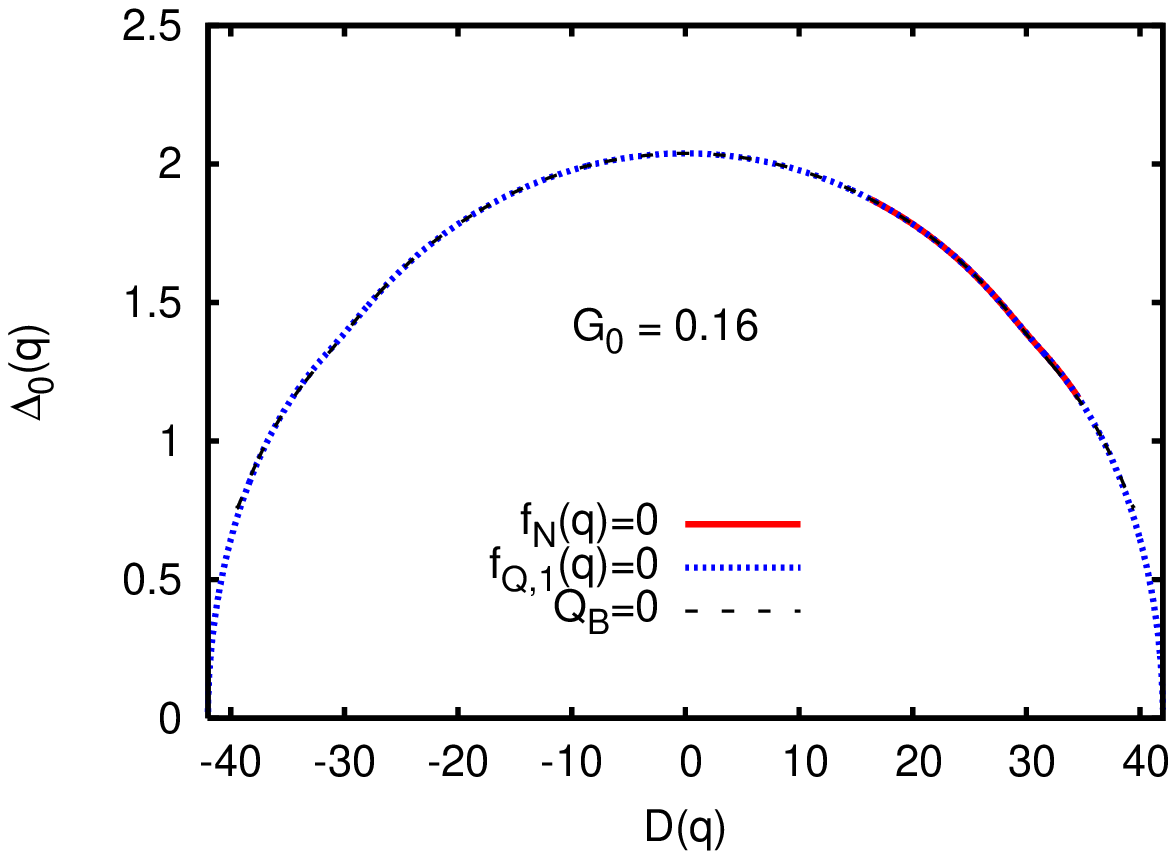} \\
\includegraphics[width=65mm]{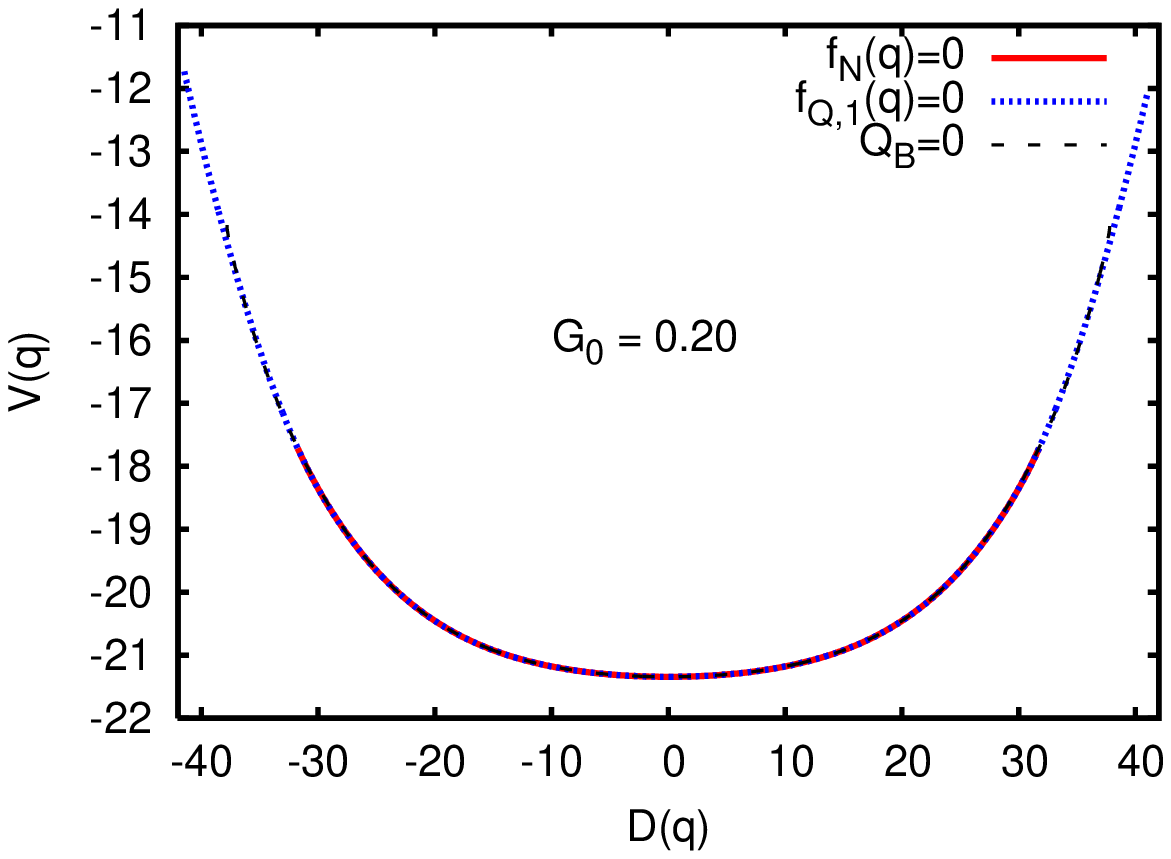} &
\includegraphics[width=65mm]{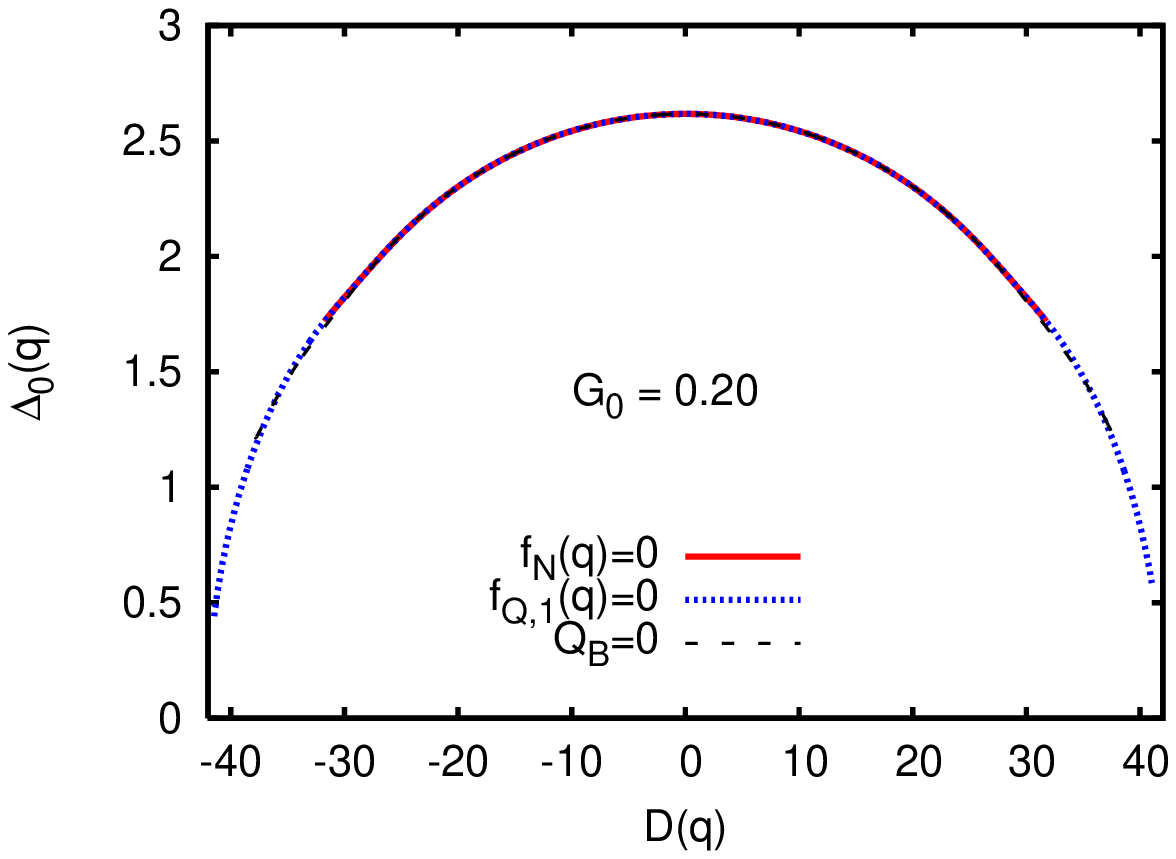}
\end{tabular}
\end{center}
\caption{Collective potentials $V(q)$ and 
monopole pairing gaps $\Delta_0(q)$
plotted as functions of the quadrupole deformation $D$.
The upper, middle and lower panels display the results for
$G_0=0.14, 0.16$ and $0.20$, respectively.
In each column, results of different calculation are compared;
those obtaind using the QRPA gauge ($f_N(q)=0$) and 
the ETOP gauge ($\fm_{Q,1}(q)=0$) are plotted by
solid (red) and dotted (blue) lines, respectively, while those obtaind 
ignoring the $B$-part of $\Qhat(q)$ ({\it i.e.}, putting $Q_i^{B}(q)=0$) 
by dashed lines.}
\label{fig:VD0}
\end{figure}

\begin{figure}[htbp]
\begin{center}
\begin{tabular}{cc}
\includegraphics[width=65mm]{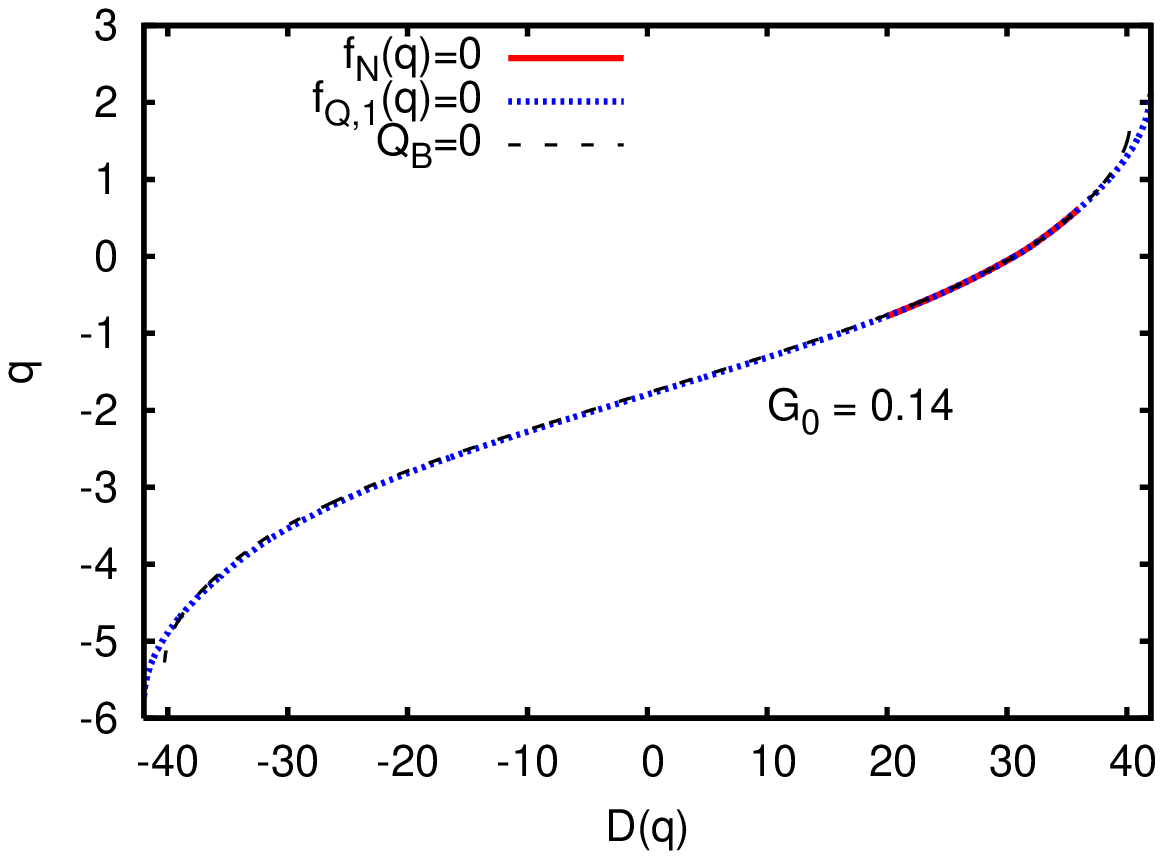} &
\includegraphics[width=65mm]{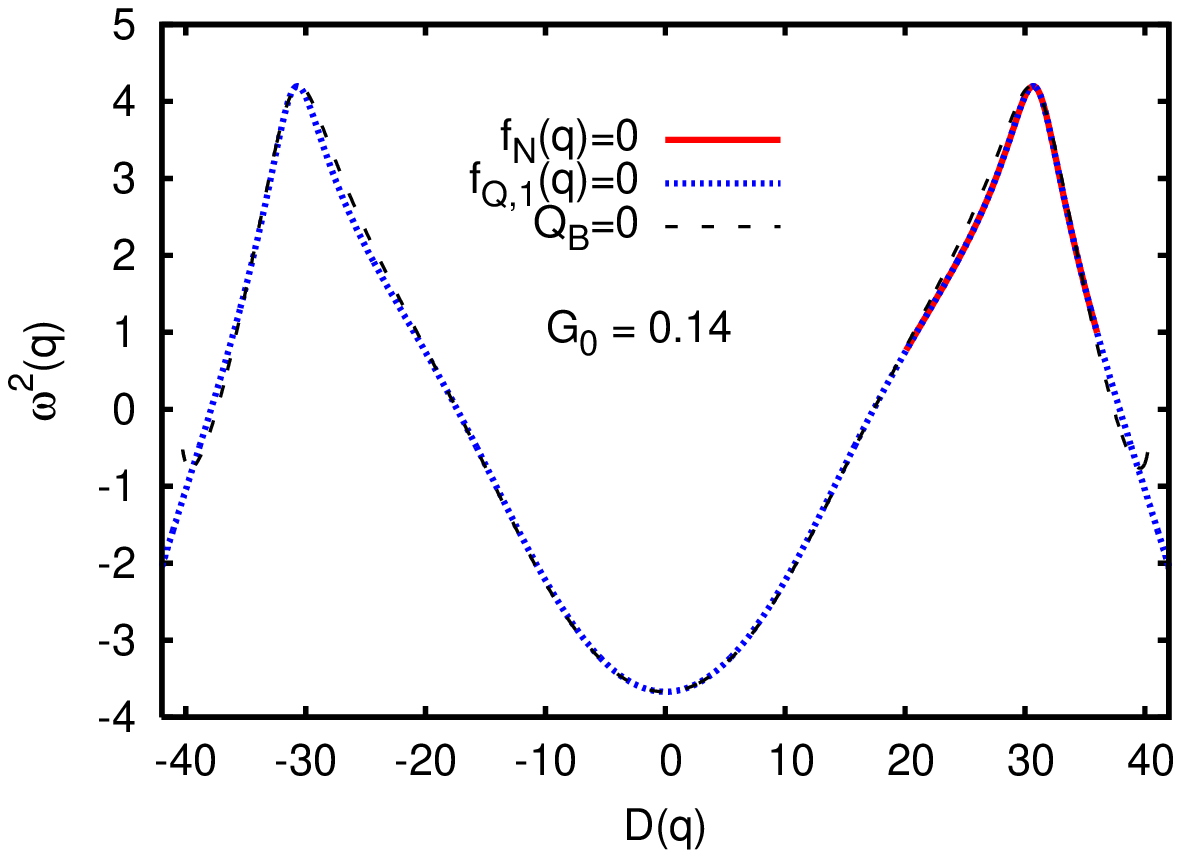} \\
\includegraphics[width=65mm]{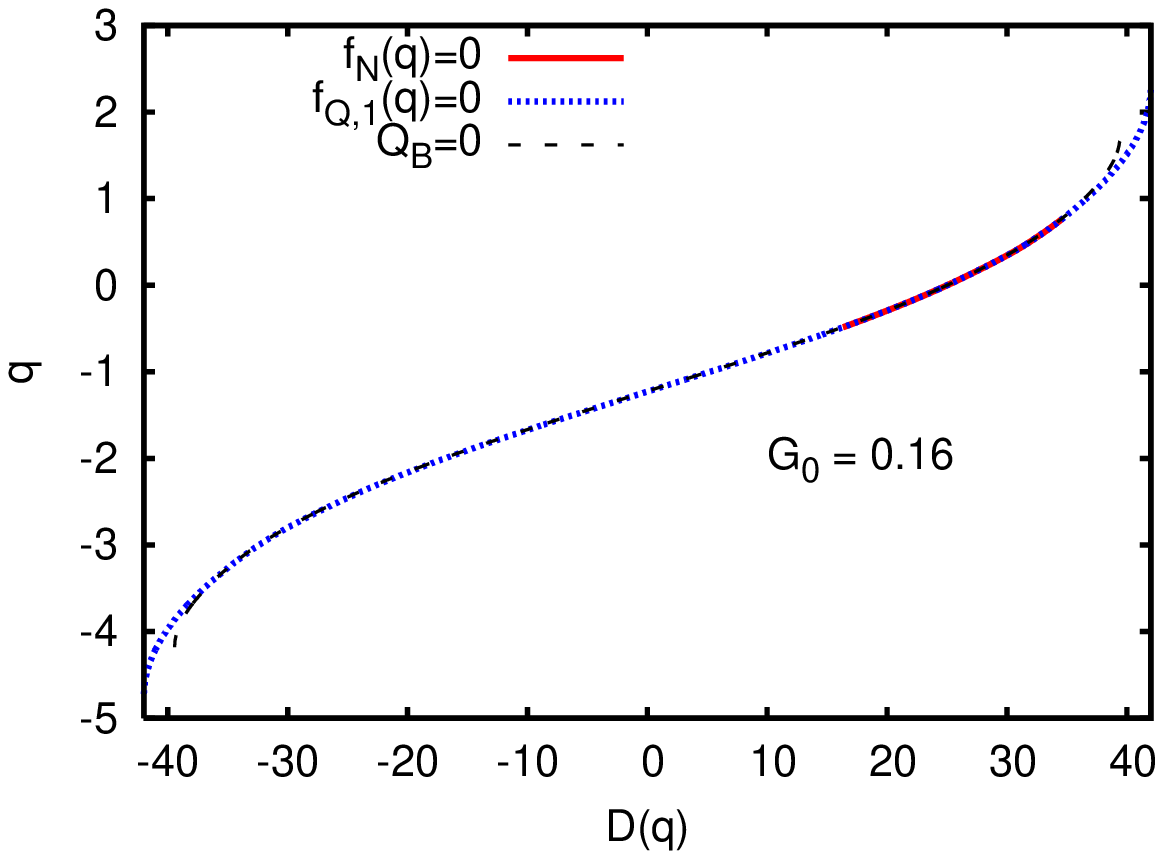} &
\includegraphics[width=65mm]{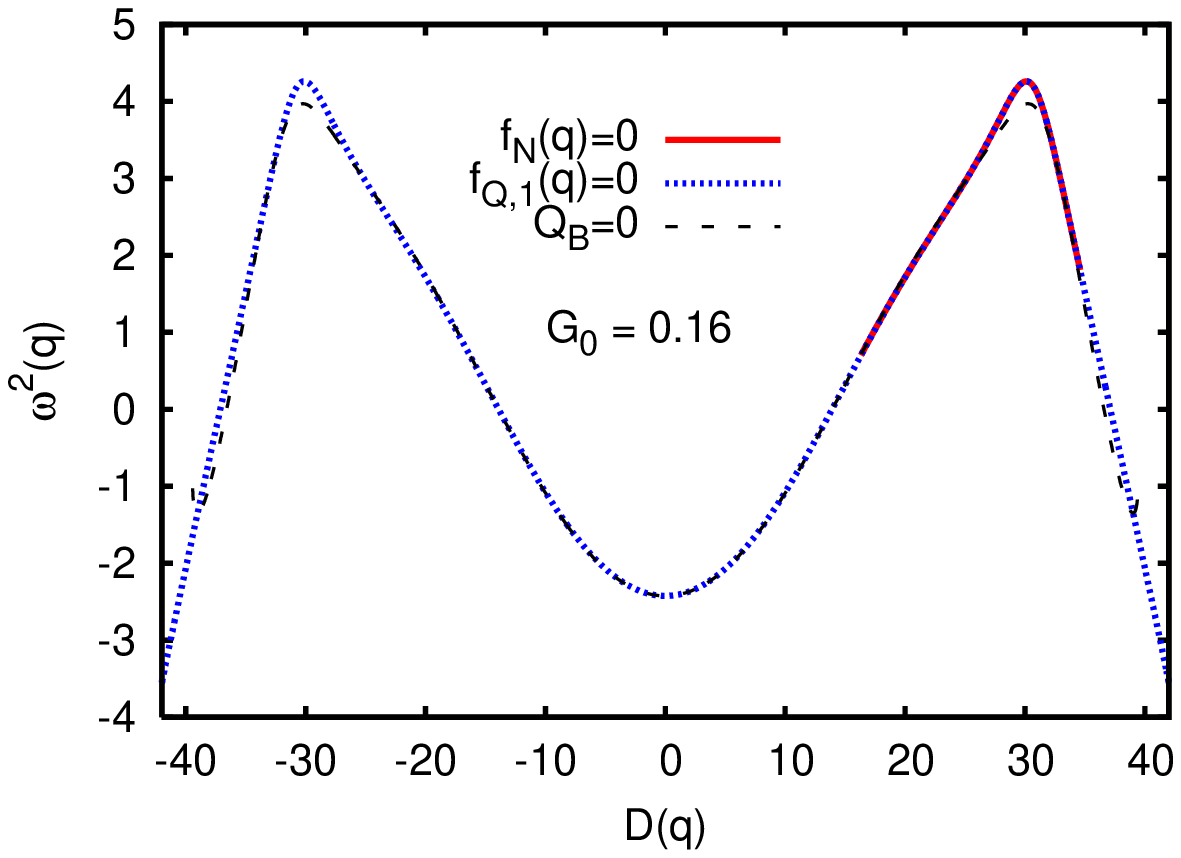} \\
\includegraphics[width=65mm]{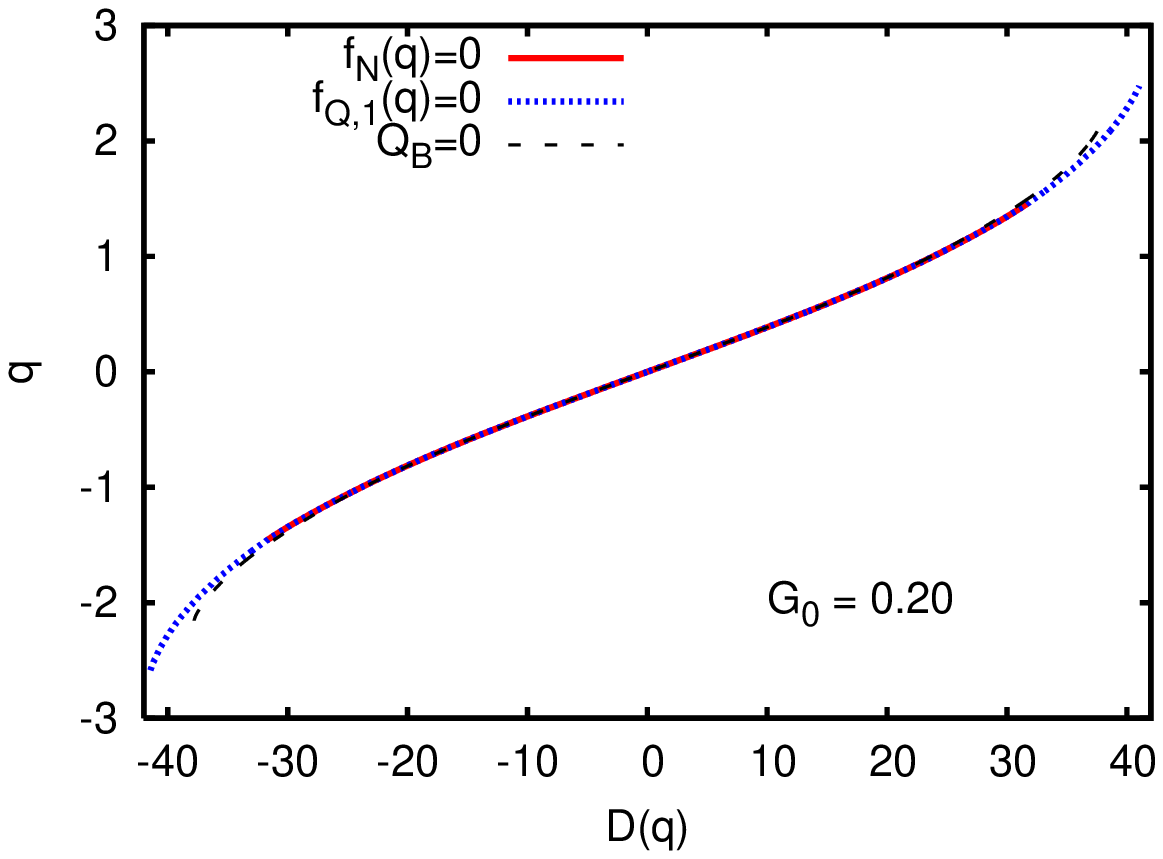} &
\includegraphics[width=65mm]{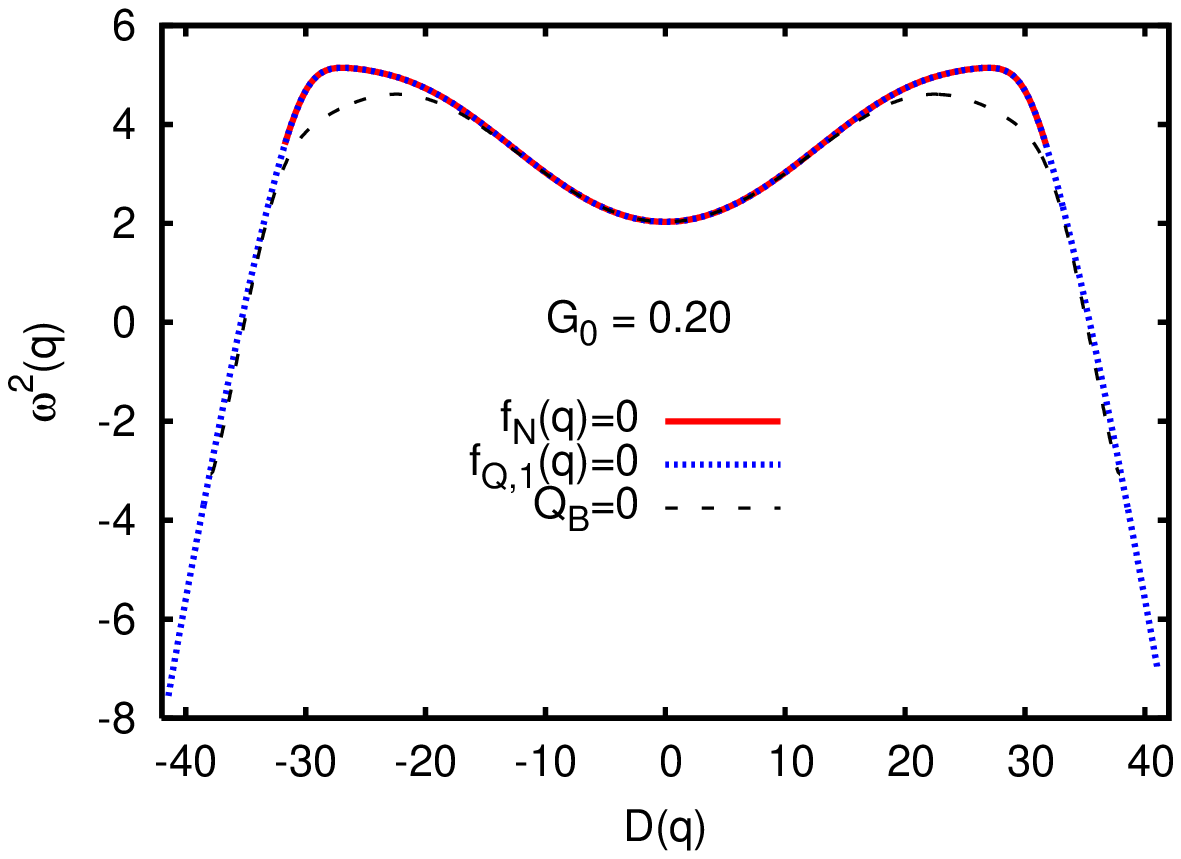} 
\end{tabular}
\end{center}
\caption{
{\it Left column}:
Relation between the collective coordinate $q$ 
and the quadrupole deformation $D(q) = \bra{\phi(q)}\Dhat\ket{\phi(q)}$.
The point $q=0$ corresponds to the HFB equilibrium, 
which is the starting point of the numerical calculation. 
{\it Right column}:
Squared frequencies $\omega^2(q)$ of the local harmonic equations,
plotted as functions of $D$.
Note that they are negative; i.e., $\omega(q)$ is purely imaginary
in the region where the curvature of the collective potential is negative.
The upper, middle and lower rows display the results for
$G_0=0.14, 0.16$ and $0.20$, respectively.
See the caption of Fig.~\ref{fig:VD0}.
}
\label{fig:q-D}
\end{figure}

\begin{figure}[htbp]
\begin{center}
\begin{tabular}{c}
\includegraphics[width=100mm]{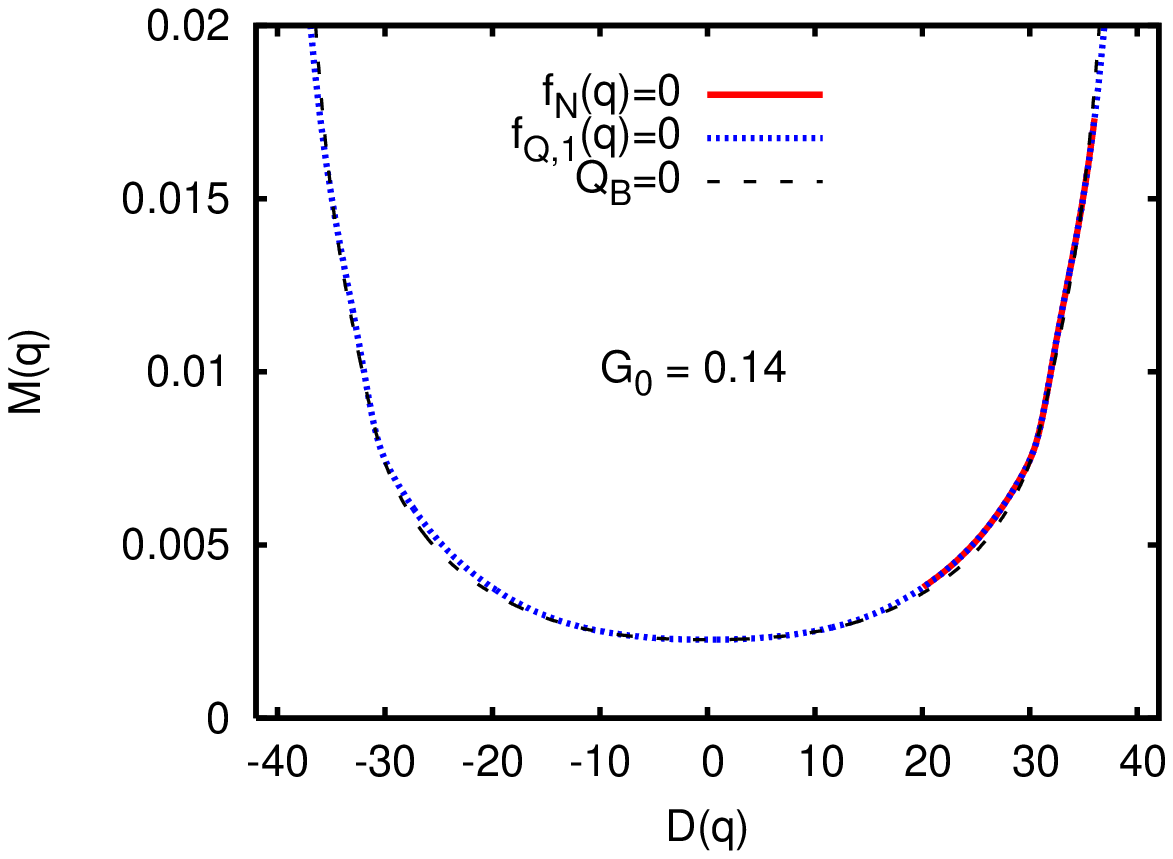} \\
\includegraphics[width=100mm]{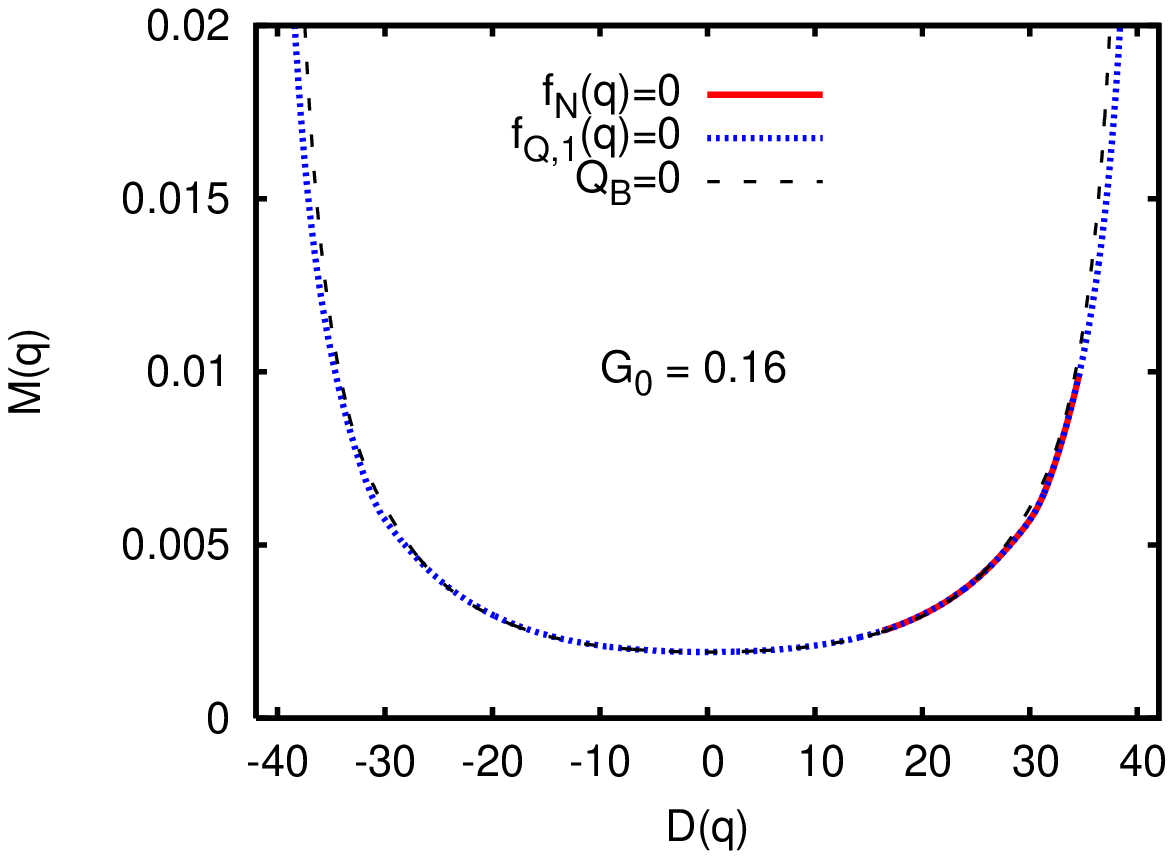} \\
\includegraphics[width=100mm]{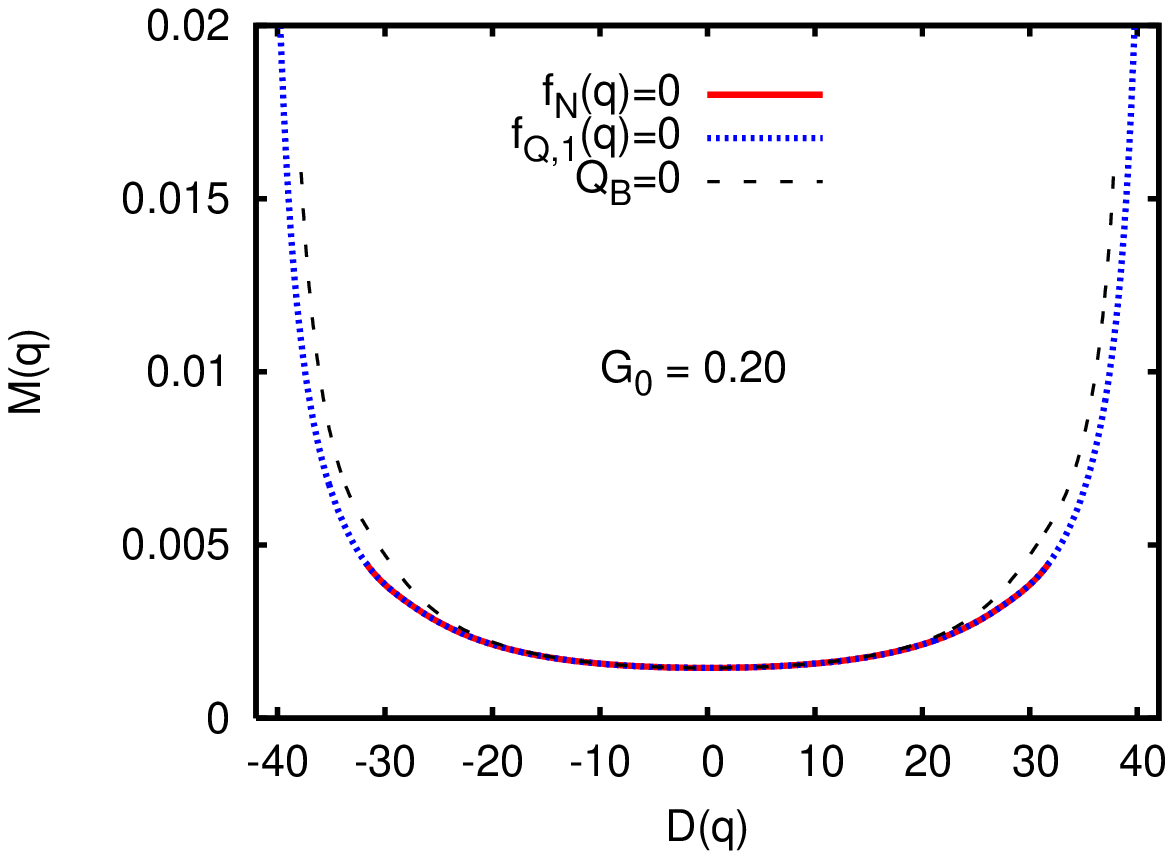} 
\end{tabular}
\end{center}
\caption{
The ASCC collective mass $M(q(D))$
as functions of the deformation $D$.
The upper, middle and lower rows display the results for
$G_0$ = 0.14, 0.16 and 0.20, respectively.
See the caption of Fig.~\ref{fig:VD0}.}
\label{fig:Q20}
\end{figure}

Figure~\ref{fig:VD0} shows the collective potential $V(q)$ and 
the monopole pairing gap $\Delta_0(q)$ as functions of the quadrupole 
deformation $D(q)$. 
Figure~\ref{fig:q-D} displays the relation between the collective variables 
$q$ and the quadrupole deformation $D$ as well as 
the squared frequency $\omega^2(q)$ obtained 
by solving the local-harmonic equations. 
The collective mass $M(D(q))=(dq/dD)^2$ is plotted as a function of $D$ 
in Fig.~\ref{fig:Q20}. 
We find that the calculation using the ETOP gauge works very well
and the collective path connecting the two local (oblate and prolate) minima
with different signs of deformation are successfully obtained. 
In contrast, the calculation using the QRPA gauge
encounters a point where we cannot proceed any more.
In the region where the solutions have been found for both gauges,
they well agree with each other.  It should be the case because 
these are gauge invariant quantities.
The cause of the difficulty encountered in the QRPA gauge is 
understood in the following analysis.

\begin{figure}[htbp]
\begin{center}
\begin{tabular}{c}
\includegraphics[width=80mm]{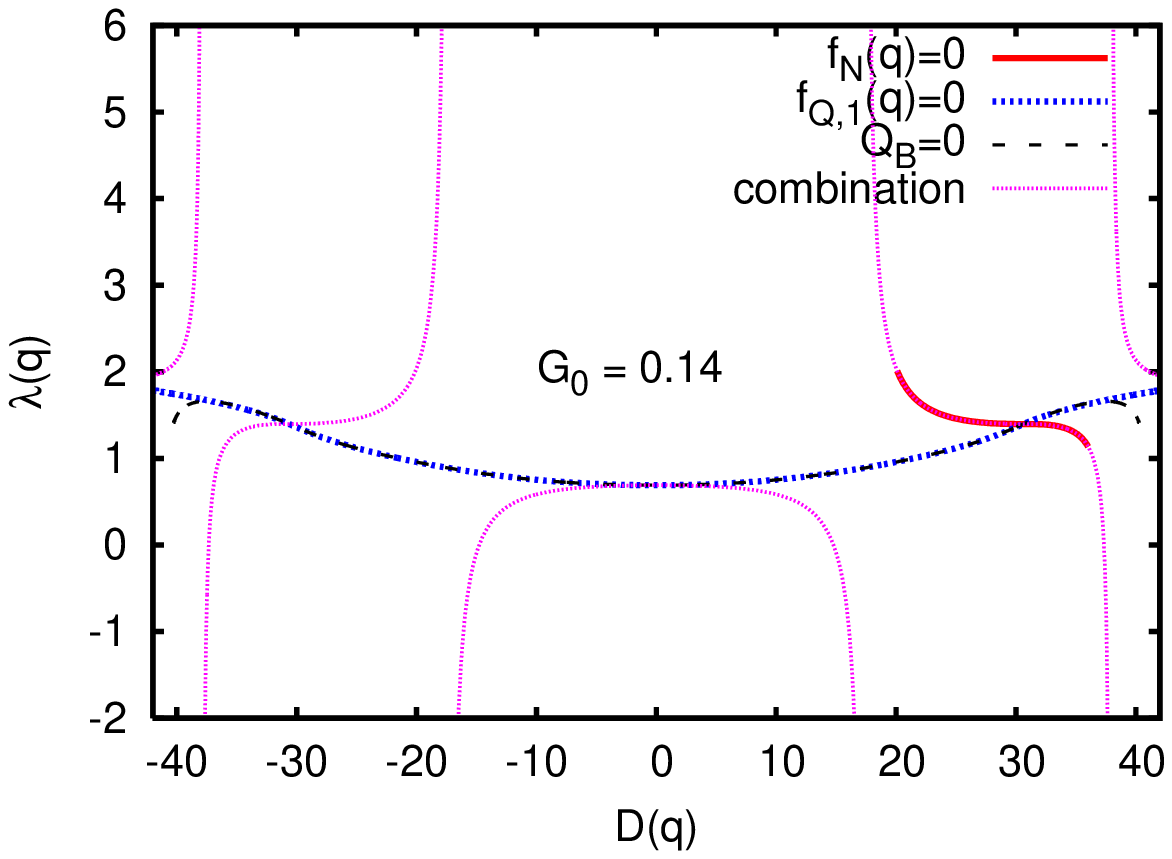} \\
\includegraphics[width=80mm]{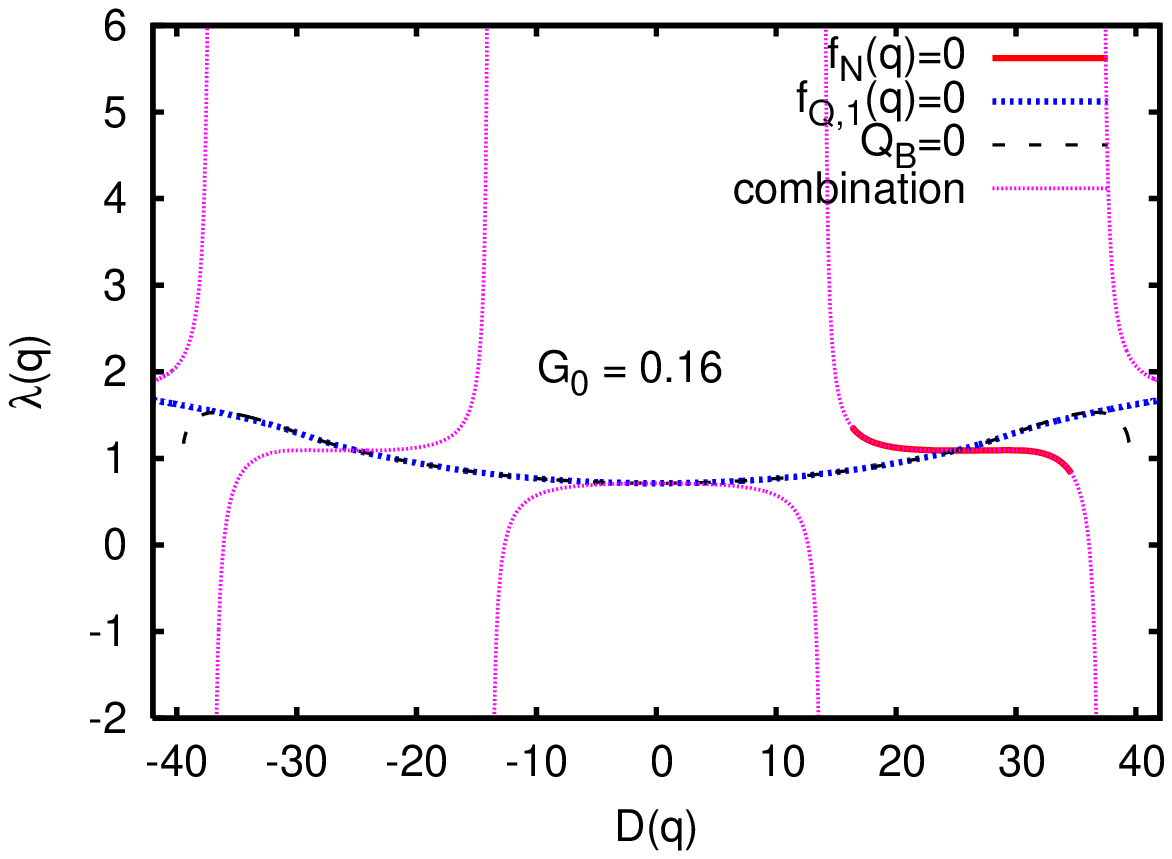} \\
\includegraphics[width=80mm]{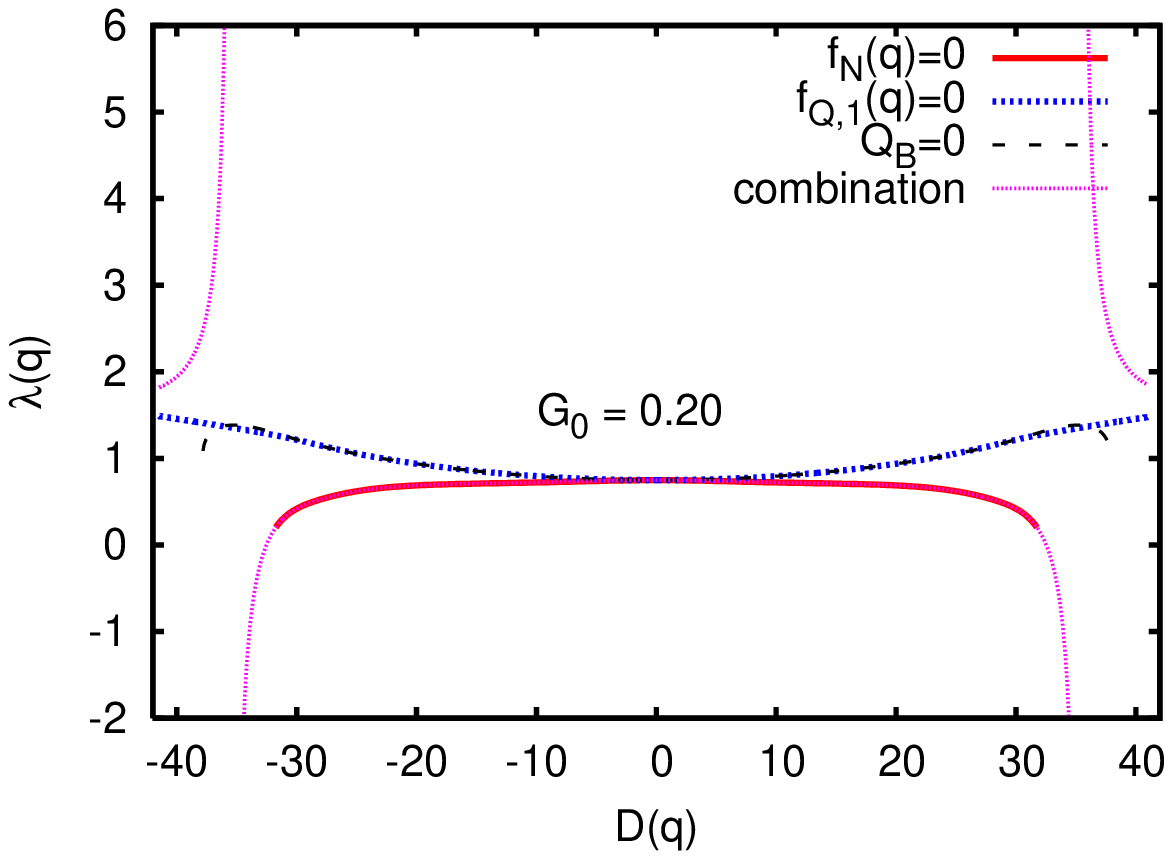}
\end{tabular}
\end{center}
\caption{
The chemical potentials $\lambda(q)$
as functions of the deformation $D$.
The upper, middle and lower rows display the results for
$G_0$ = 0.14, 0.16 and 0.20, respectively.
In each column, results of different calculation are compared;
The solid (red) and dotted (blue) lines indicate the results 
obtaind using the QRPA gauge ($f_N(q)=0$) and 
the ETOP gauge ($\fm_{Q,1}(q)=0$), respectively, 
while the dashed lines indicate those obtaind using the ETOP gauge but
ignoring the $B$-part of $\Qhat(q)$ ({\it i.e.}, putting $Q_i^{B}(q)=0$). 
The dotted (purple) lines labeled ``combination''
indicate the results calculated by switching to the QRPA gauge  
after the collective path is determined by using the ETOP gauge.
}
\label{fig:lambda}
\end{figure}

\begin{figure}[htbp]
\begin{center}
\begin{tabular}{cc}
\includegraphics[width=65mm]{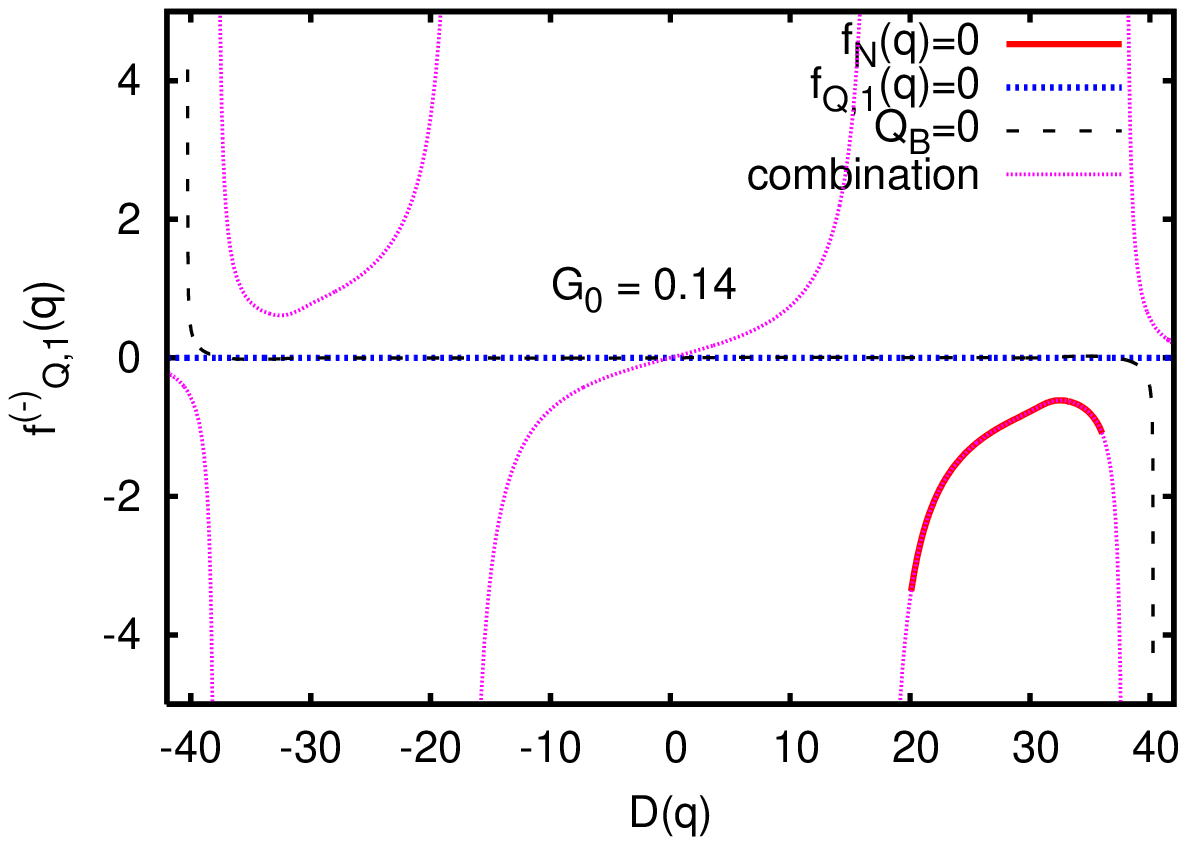} &
\includegraphics[width=65mm]{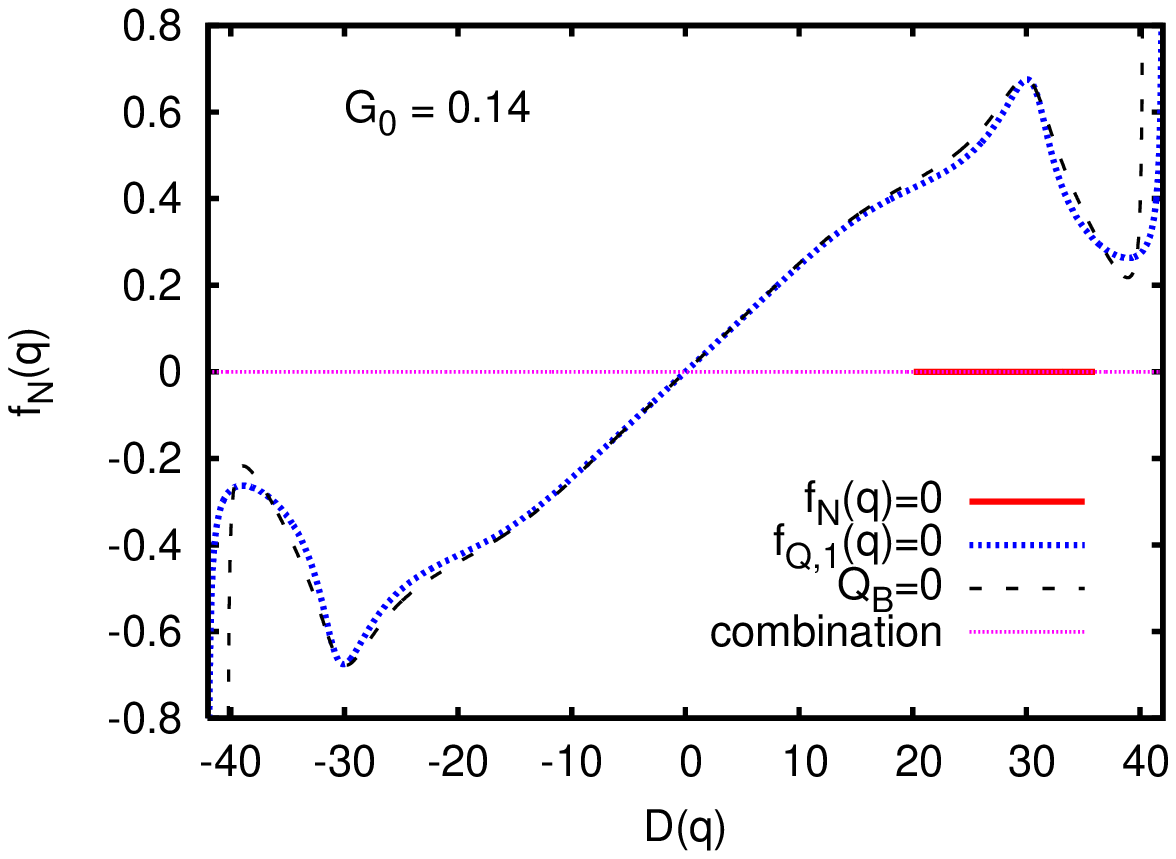} \\
\includegraphics[width=65mm]{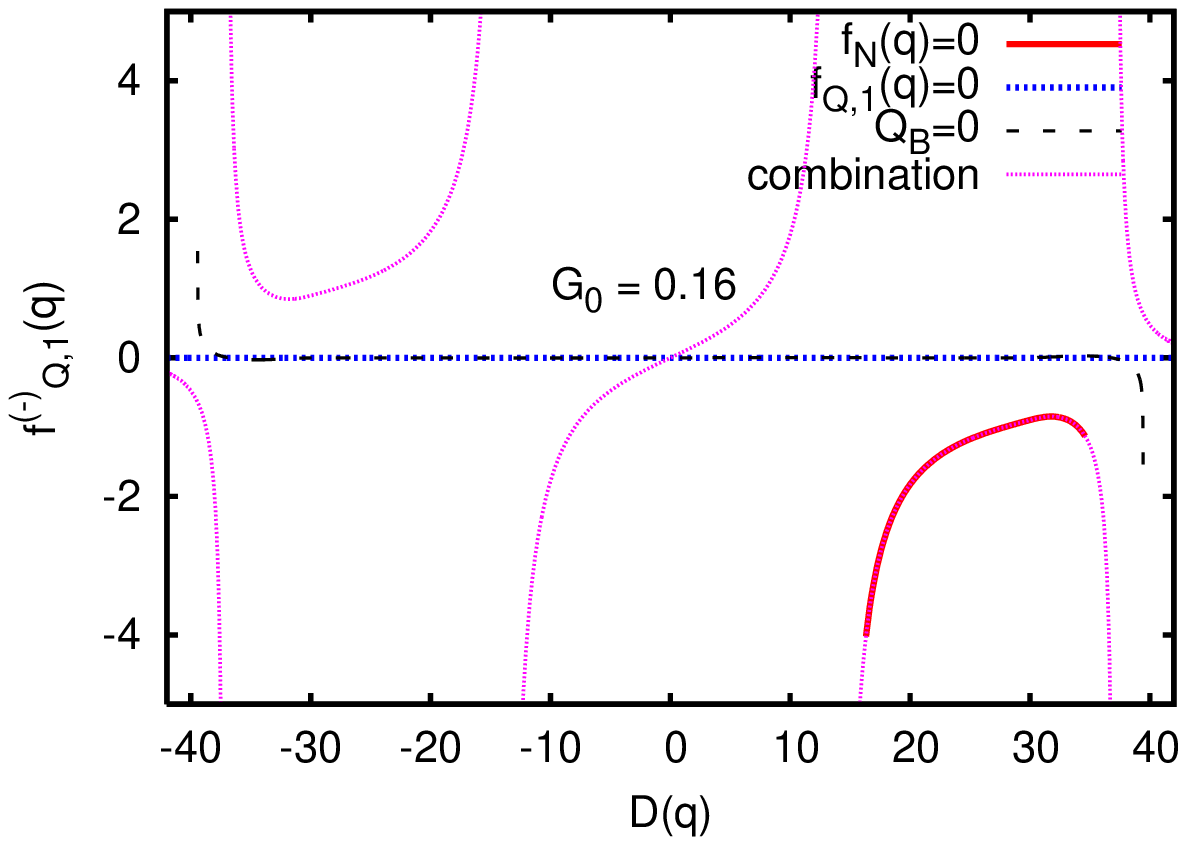} &
\includegraphics[width=65mm]{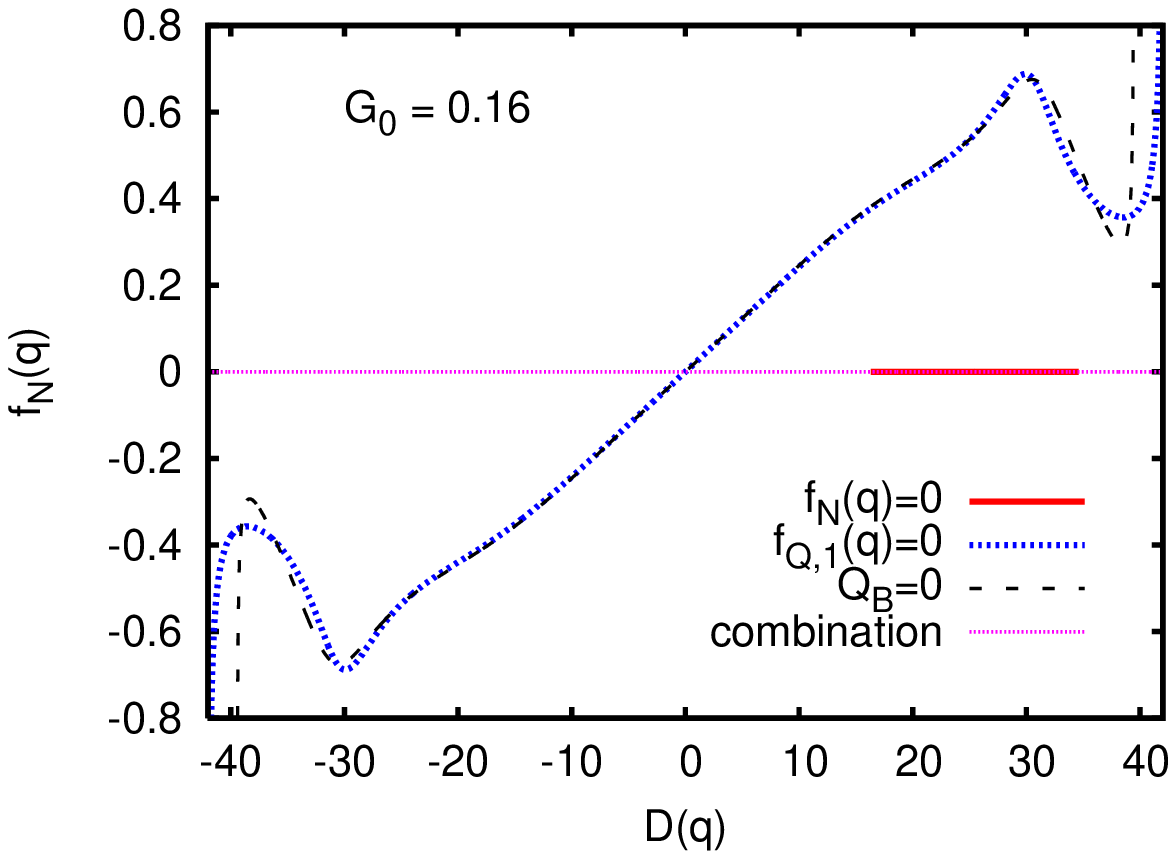} \\
\includegraphics[width=65mm]{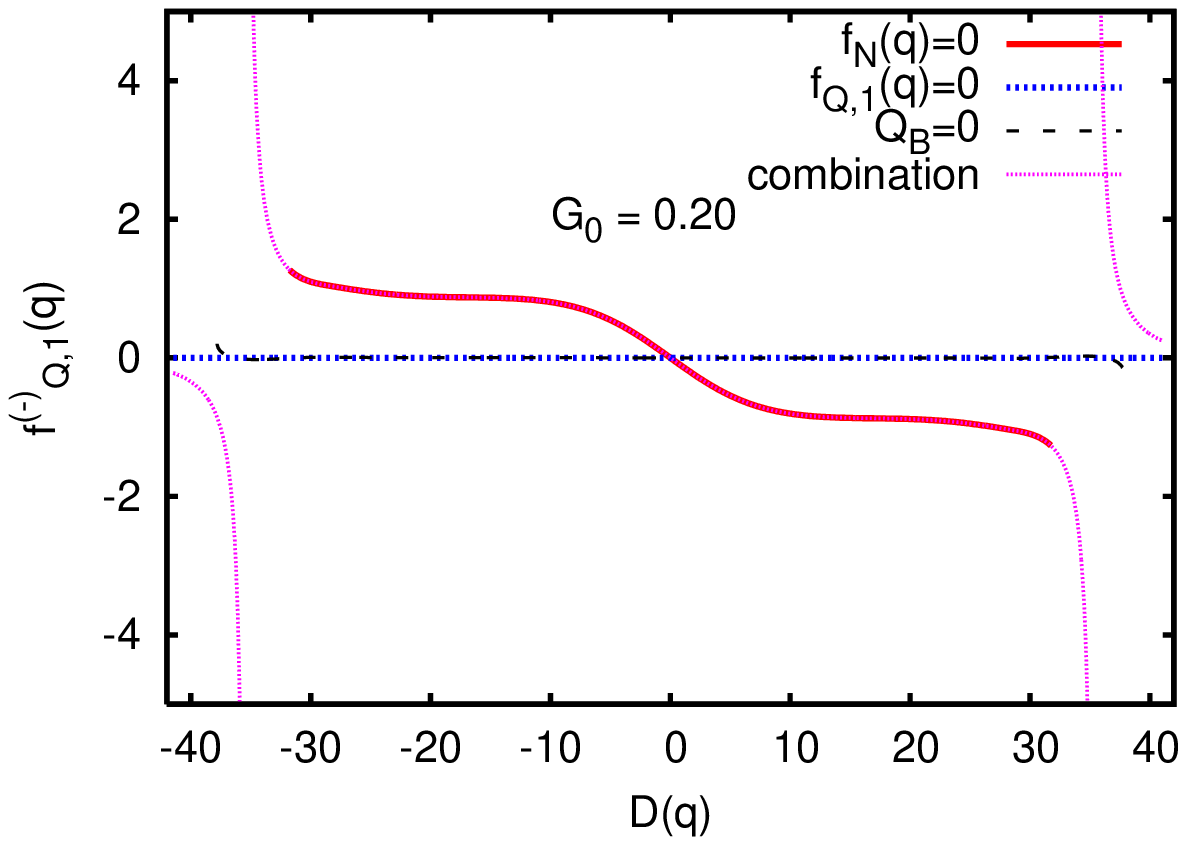} &
\includegraphics[width=65mm]{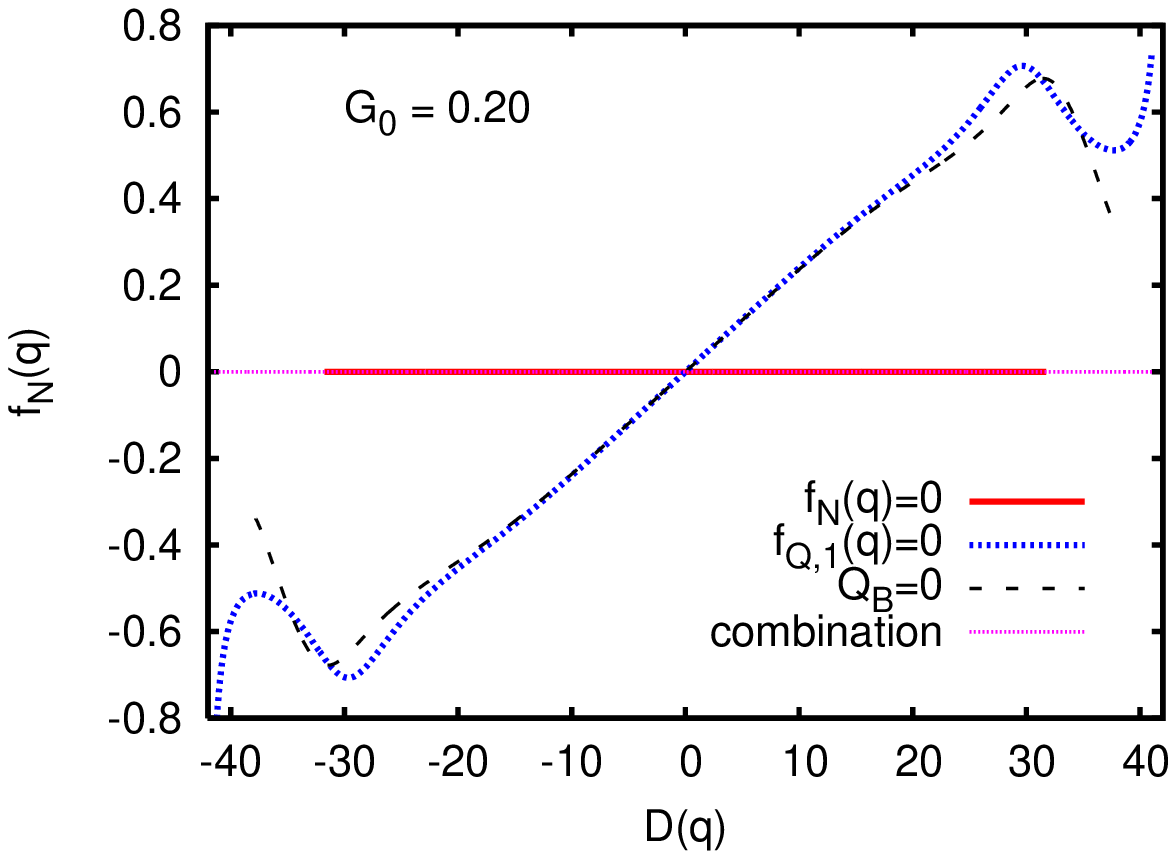}
\end{tabular}
\end{center}
\caption{Gauge dependent quantities, $\fm_{Q,1}(q)$ ({\it left column})
and $f_N(q)$ ({\it right column}), plotted as functions of 
the deformation $D$.
The upper, middle and lower rows display the results for
$G_0=0.14, 0.16$ and $0.20$, respectively.
See the caption of Fig.~\ref{fig:lambda}.}
\label{fig:f_Qf_N}
\end{figure}

Figures~\ref{fig:lambda} and \ref{fig:f_Qf_N} display 
the chemical potential $\lambda(q)$ 
and the quantities, $\fm_{Q,1}(q)$ and $f_N(q)$, respectively.
Their values depend on the gauge adopted.
If the QRPA gauge is used,  $\lambda(q)$ should be constant along
the path because of the condition $f_N(q)=B(q)\del\lambda/\del q=0$.
We find, however, $\lambda(q)$ diverges near the inflection point 
of the collective potential where $\omega(q)^2=\del^2V/\del q^2=0$. 
This divergence occurs because the inflection point is a singularity 
for the gauge transformation (\ref{eq:gauge-f_N}) where an arbitrary 
$\alpha$ gives the same $f_N(q)$. 
Thus, the calculation using the QRPA gauge stops at the inflection point.
On the other hand, we can go over the inflection point 
using the ETOP gauge, because the gauge transformation 
for $\fm_{Q,1}(q)$ (\ref{eq:gauge-f_Q1}) involves only 
the monopole pairing gap $\Delta_0(q)$
which always takes finite values along the collective path
(except at the limit of the model space).
In these figures we also present the results that are obtained 
by the following procedure: 
After determining the collective paths with the use of the ETOP gauge, 
we calculate the gauge-dependent quantities,
$\lambda(q), f_{Q,1}^{(-)}(q)$, $f_{N}(q)$, 
by switching to the QRPA gauge using the relations
\begin{subequations} \label{eq:transETOP}
\begin{align}
 \fm_{Q,1 ({\rm QRPA})} =& -\frac{ 4f_N(q)_{({\rm
 ETOP})}\Delta_0(q)}{\omega^2(q)} \label{eq:transETOP-fQ1}\\
 \lambda(q)_{({\rm QRPA})} =& \lambda(q)_{({\rm ETOP})}
 - \frac{4f_N(q)_{({\rm ETOP})}}{\omega^2(q)} \frac{\del V}{\del q}
 \label{eq:transETOP-lambda}
\end{align}
\end{subequations}
We see in Figs.~\ref{fig:lambda} and \ref{fig:f_Qf_N} that
the results obtained by this procedure agree with those
calculated by using the QRPA gauge ($f_N(q)=0$) from the beginning
(in the region of deformation $D(q)$ where the collective path can be 
obtained using the QRPA gauge).
This agreement demonstrates that the collective paths determined by 
using different gauge fixing conditions are the same, as it should be. 
Nevertheless, there is a suitable gauge fixing condition 
in finding solutions of the ASCC equations and constructing the collective path.
For the multi-$O$(4) model with superfluidity, 
we find that the ETOP gauge is more useful than the QRPA gauge, 
because the gauge transformation (\ref{eq:gauge-f_Q1})
is well defined as long as the pairing gap $\Delta_0(q)$ is non-zero.

\subsection{Comparison with the previous calculation}

In our previous paper\cite{hin06}, we employed 
the ETOP gauge condition ($\fm_{Q,1}(q)=0$) but the
$B$-part of $\Qhat(q)$ is ignored.
Let us evaluate the error caused by this approximation.
The results of such calculation are presented also in Figs.~\ref{fig:VD0}-\ref{fig:lambda}
and compared with those of the full calculation.
We see that they agree well indicating that the approximation 
of ignoring the $B$-part is rather good.

In Fig.~\ref{fig:f_Qf_N}, we present the quantity 
\begin{align}
\fm_{Q,1}(q) = -\kappa_s \bra{\phi(q)}[\Fhatm_s, \Qhat(q)]\ket{\phi(q)}
= -\kappa_s \sum_{i>0} Q^A_i(q) \label{eq:checkfQ1}
\end{align}
evaluated using the $\Qhat(q)$ operator
that is obtained by ignoring the $B$-part in the process of 
solving the ASCC equations.
This quantity should be zero if the $\Qhat(q)$ operator determined by 
the gauge-invariant ASCC equations is used.
We see that the deviation from zero is negligible 
(except near the limit of the model space),
again indicating that the approximation is good.

\begin{figure}[htbp]
\begin{center}
\begin{tabular}{cc}
\includegraphics[width=65mm]{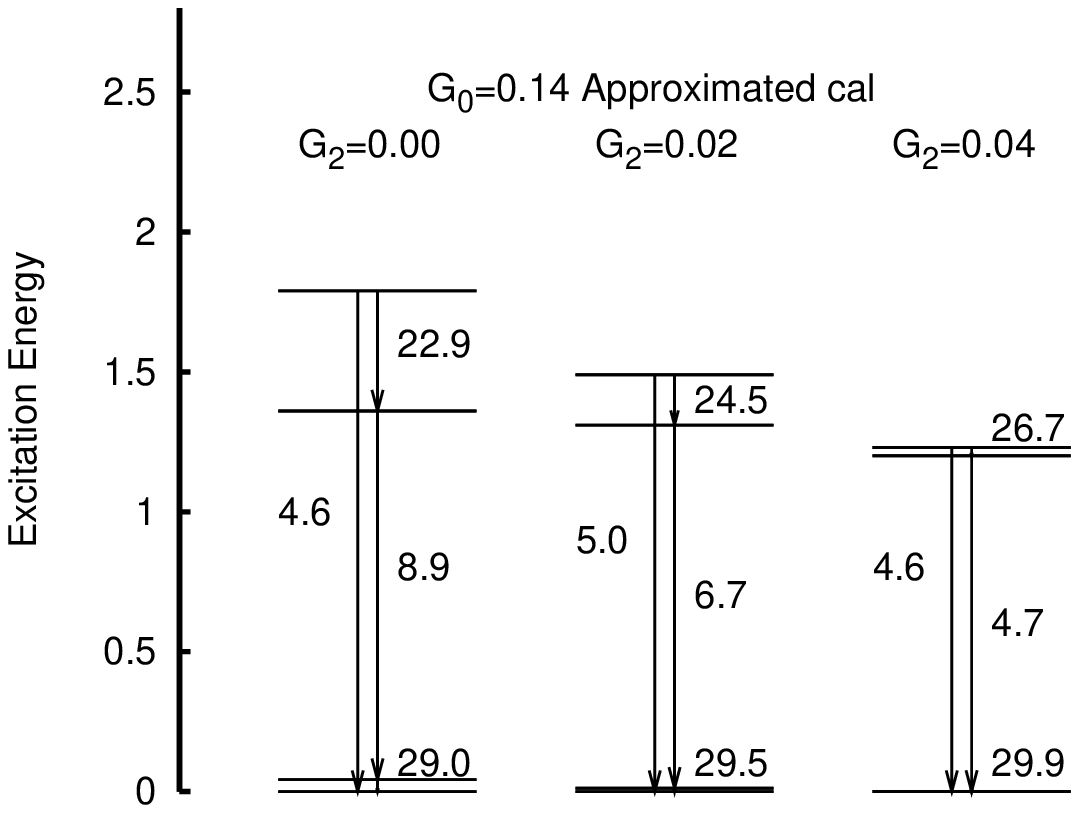} &
\includegraphics[width=65mm]{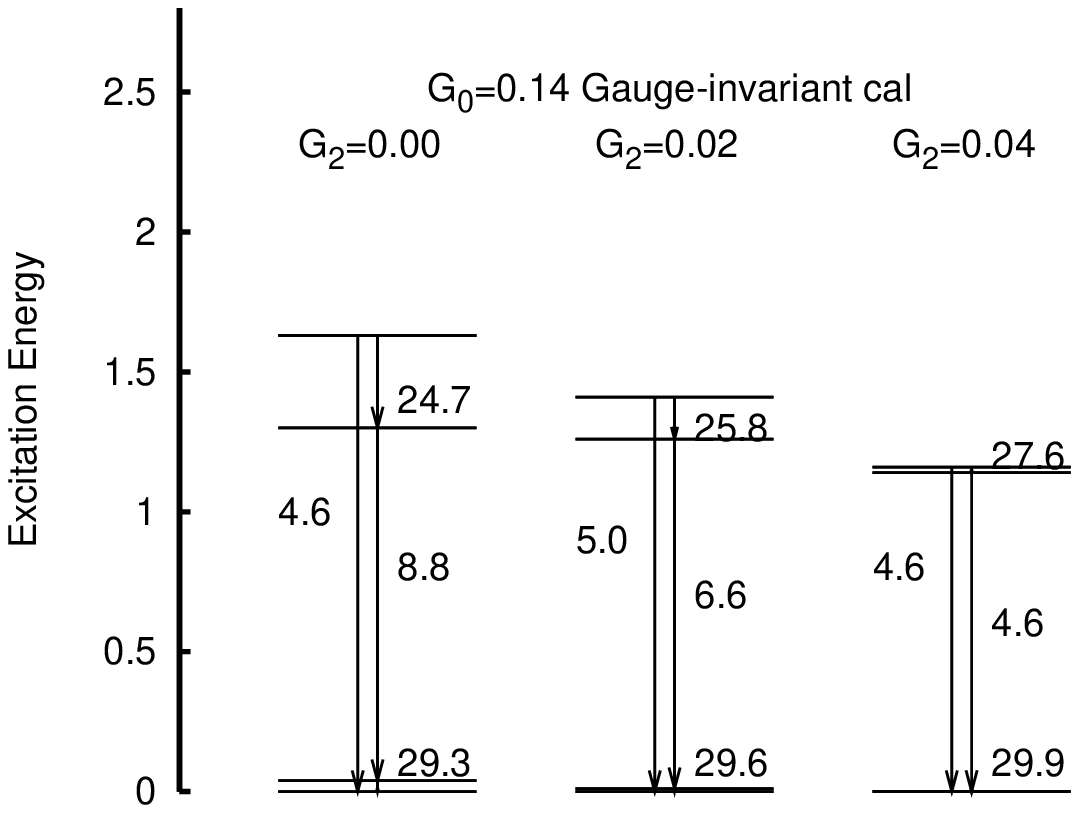} \\
\includegraphics[width=65mm]{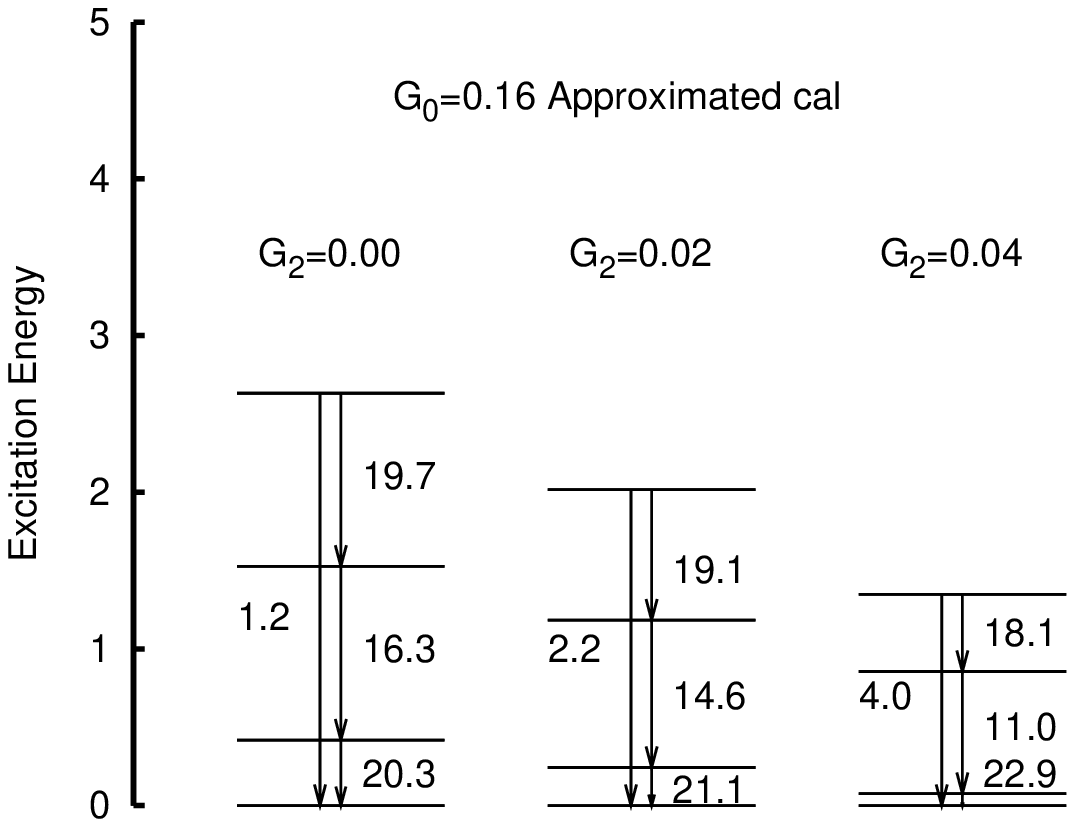} &
\includegraphics[width=65mm]{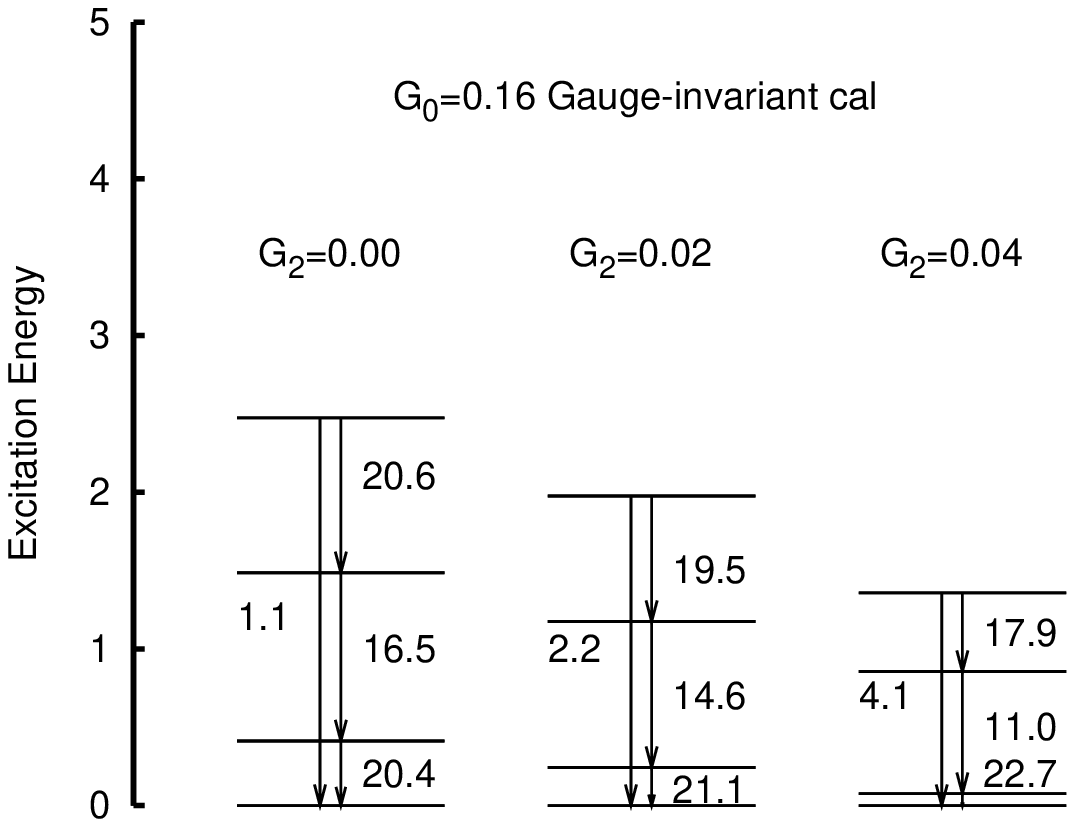} \\
\includegraphics[width=65mm]{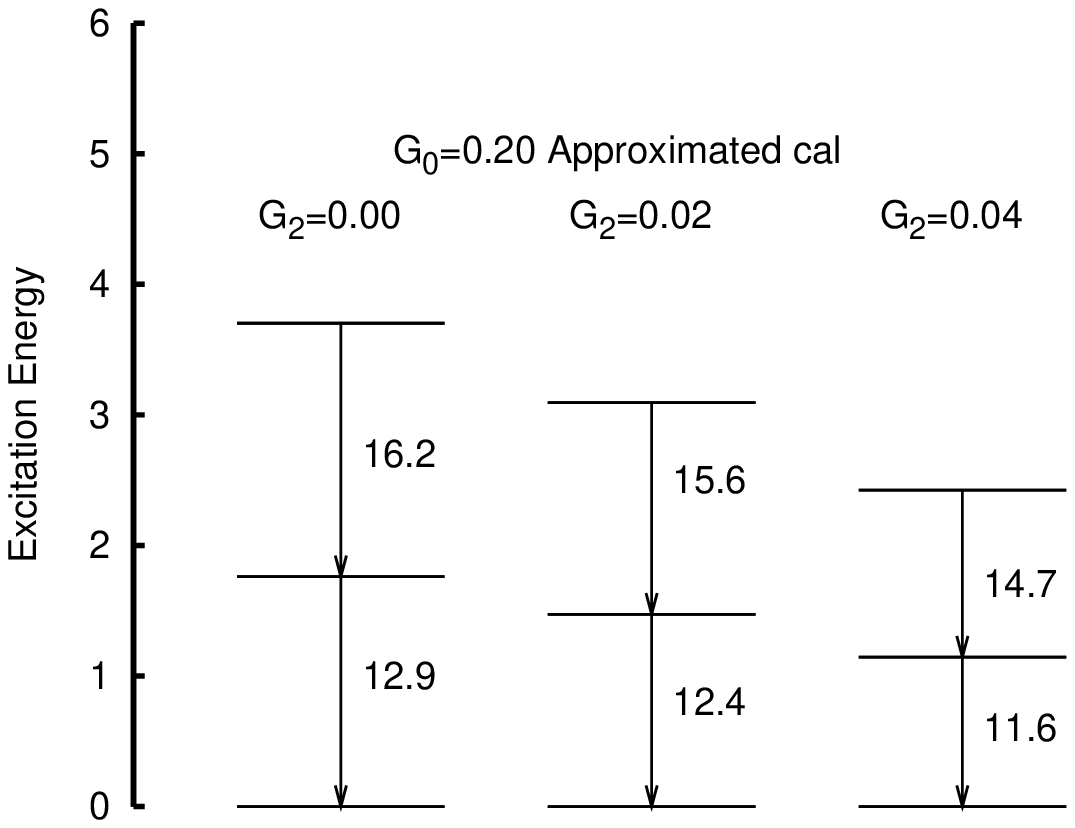} &
\includegraphics[width=65mm]{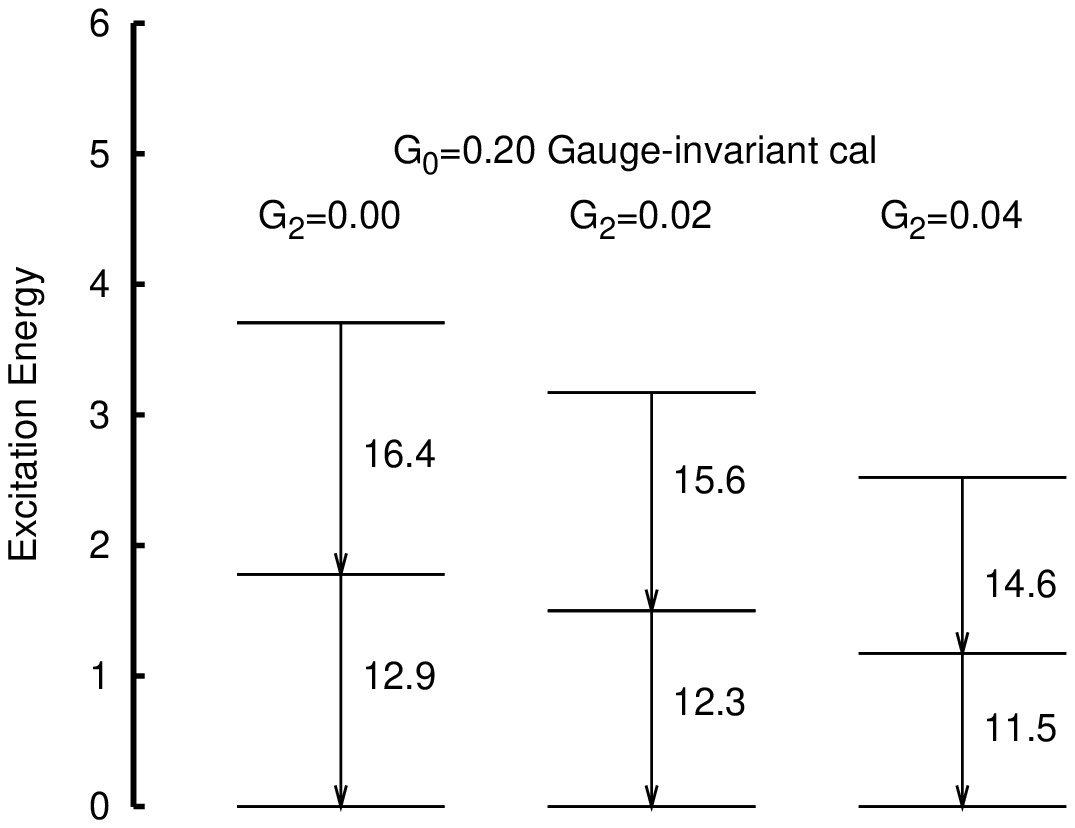} 
\end{tabular}
\end{center}
\caption{Excitation spectra and quadrupole transition matrix elements.
{\it left column}: 
The result obtaind using the ETOP gauge ($\fm_{Q,1}(q)=0$) but
ignoring the $B$-part of the collective coordinate operator $\Qhat(q)$.
This is the same as presented in Ref.~\citen{hin06}.~
{\it right column}: 
The result obtained by solving the gauge-invariant ASCC equations 
using the ETOP gauge.
The upper, middle and lower rows display the results for
$G_0$ = 0.14, 0.16 and 0.20, respectively.
In each row, the results for $G_2= 0.00, 0.02$ and $0.04$ are compared.
The numbers adjacent to the vertical lines are the absolute values of 
the transition matrix elements.
The matrix elements between the doublets are indicated beside them.
}
\label{fig:spectra.ascc}
\end{figure}

\begin{figure}[htbp]
\begin{center}
\begin{tabular}{c}
\includegraphics[width=95mm]{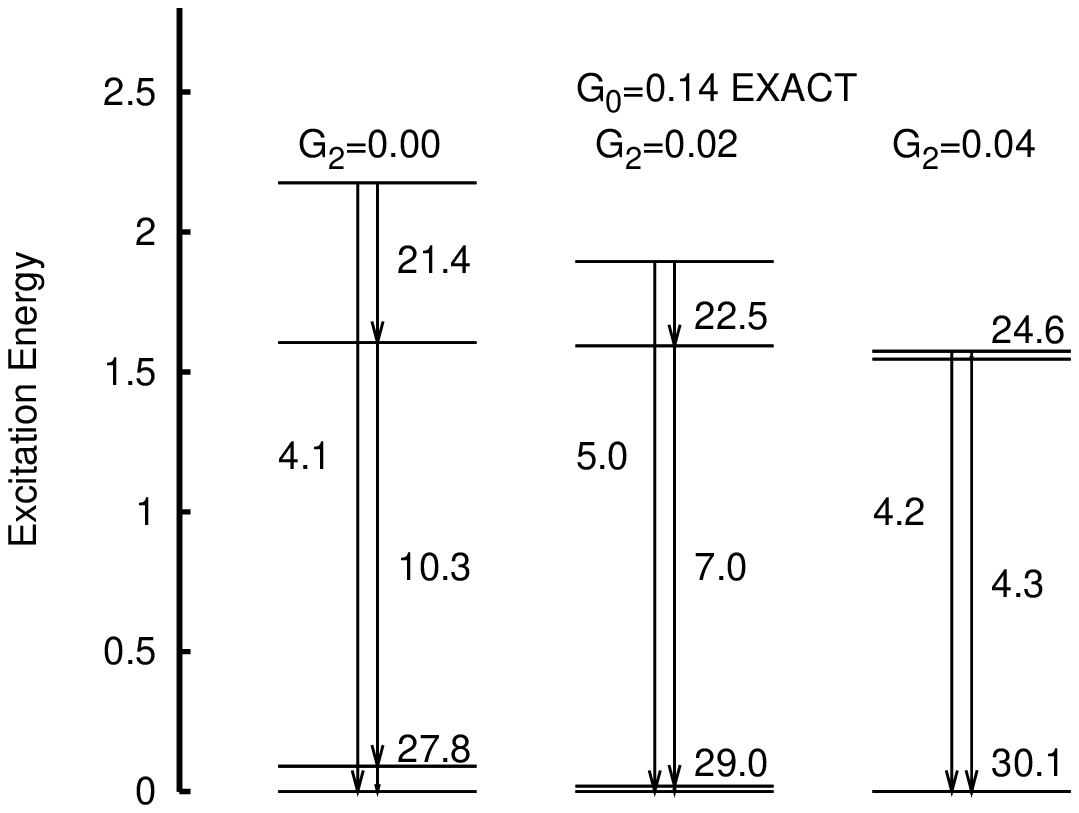} \\
\includegraphics[width=95mm]{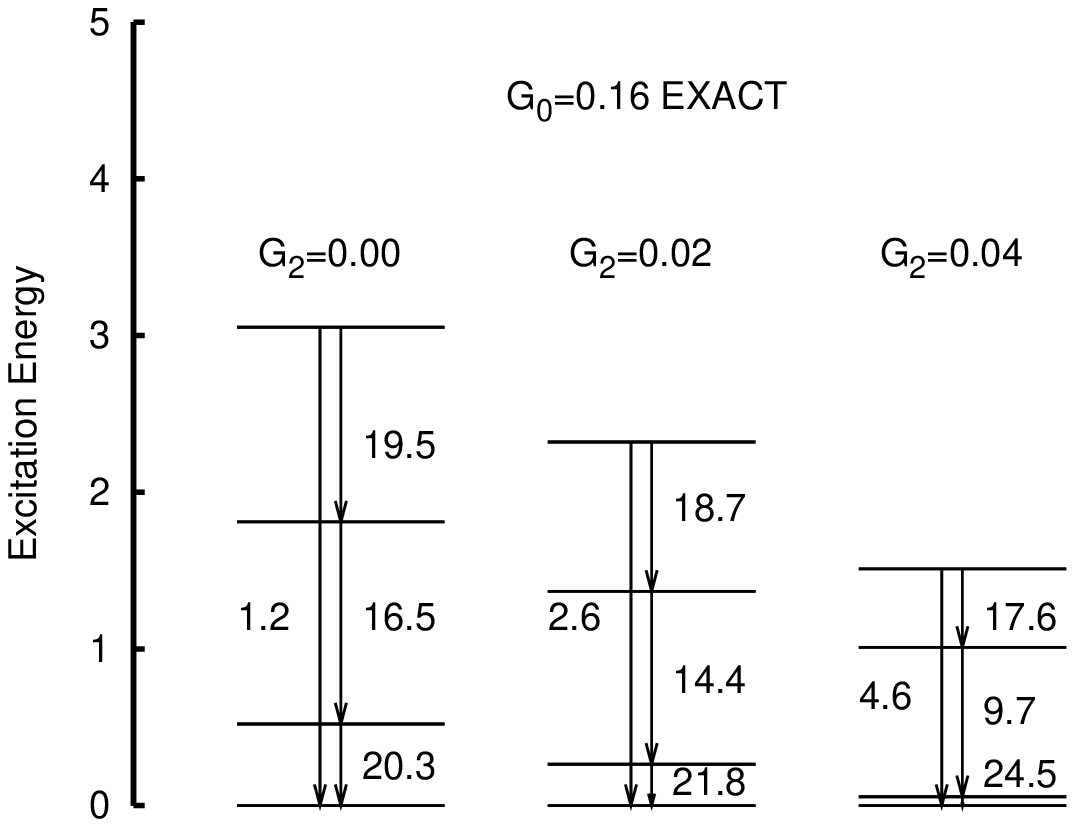} \\
\includegraphics[width=95mm]{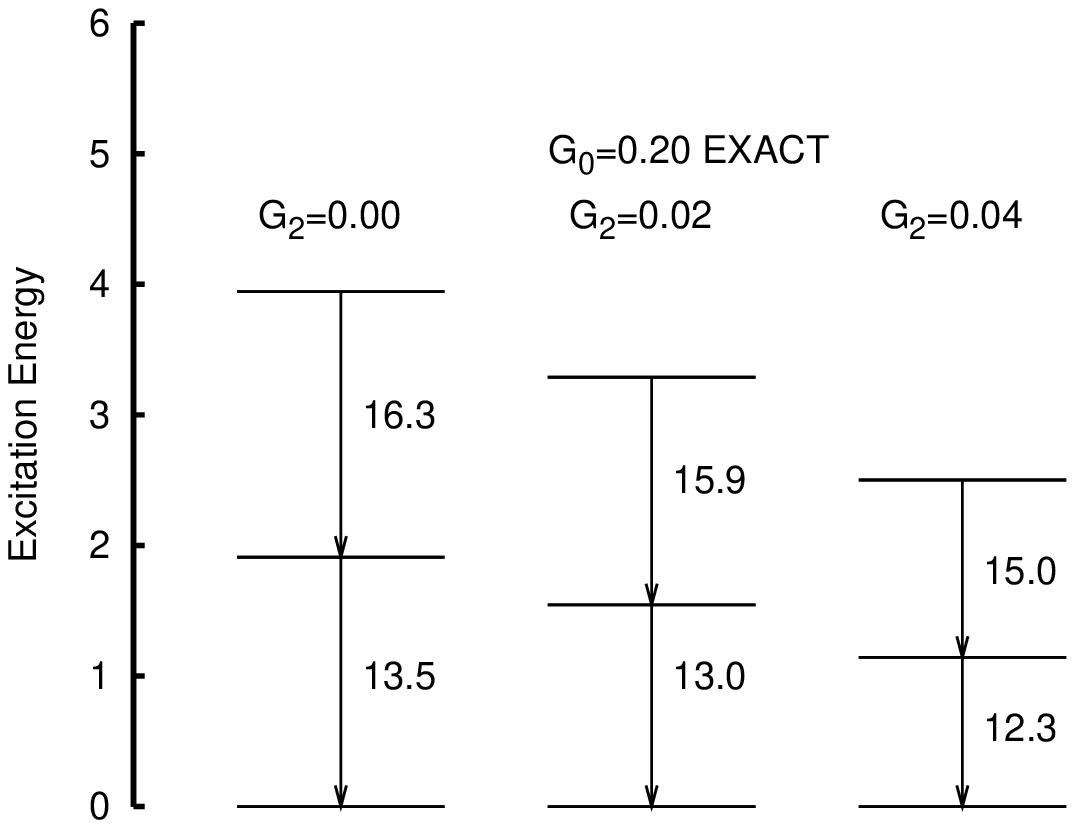}
\end{tabular}
\end{center}
\caption{Excitation spectra calculated with the exact diagonalization
of the multi-$O(4)$ model Hamiltonian.
The upper, middle and lower rows display the results for
$G_0$ = 0.14, 0.16, and 0.20, respectively.
In each row, the results for $G_2= 0.00, 0.02, 0.04$ are compared.
The numbers adjacent to vertical lines indicate the transition matrix elements.
The matrix elements between the doublets are indicated beside them.
}
\label{fig:spectra.exact}
\end{figure}

The quantum spectra and transition strengths are displayed in Fig.~\ref{fig:spectra.ascc}.
These are obtained by solving the Schr\"odinger equation for 
the quantized collective Hamiltonian.
In this figure, effects of the quadrupole pairing interaction 
are also shown. 
We see that the results of the previous calculation 
(in which the $B$-part of $\Qhat(q)$ is ignored) are quite similar 
to those of full calculation (including the $B$-part), and
both results well reproduce the trend of the excitation spectra 
obtained by exact diagonalization of the microscopic Hamiltonian
(Fig.~\ref{fig:spectra.exact}).
The numerical calculation presented above thus suggest that the
approximation of ignoring the $B$-part of $\Qhat(q)$,
adopted in our previous paper\cite{hin06}, is justified and it
may serve as an economical way of determining the collective path. 

\section{Concluding Remarks}
\label{sec:Conclusions}

We have shown that the basic equations of the ASCC method are invariant 
against transformations involving the angle in the gauge space
conjugate to the particle-number.
By virtue of this invariance, a clean separation 
between the large-amplitude collective motion and the pairing rotational 
motion can be achieved, enabling us to restore the particle-number symmetry 
broken by the HFB approximation. 
We have formulated the ASCC method explicitly in a gauge-invariant form,
then, applied it to the multi-$O$(4) model using different
gauge-fixing procedures.
The calculations 
using different gauges indeed yield the same results for 
gauge-invariant quantities, such as the collective path, the collective mass
parameter, and spectra obtained by the requantization of the collective
Hamiltonian. 
We have suggested a gauge-fixing prescription that can be used in 
realistic calculations.

The explicit gauge invariance requires a $B$-part ($a^\dagger a$ part)
of the collective coordinate operator $\Qhat(q)$.
Actually, Ref.~\citen{nak00} remarks that the separation of the
Nambu-Goldstone modes in the local harmonic formulation
requires higher-order terms in the collective coordinate.
This is consistent with the present conclusion in the gauge-invariant
formalism.
We have also demonstrated that the approximation to neglect the
$B$-part leads to results almost identical to those of the full calculation,
at least for the multi-$O(4)$ model.

We are now investigating the oblate-prolate shape coexistence phenomena
\cite{woo92,fis00,bou03} 
in nuclei around $^{68}$Se using the pairing-plus-quadrupole interactions
\cite{bar65,bar68,bes69} 
using the prescription
on the basis of the new formulation of the ASCC method 
proposed in this paper. The result will be reported in a near future.

\section*{Acknowledgements}

This work is supported by 
Grants-in-Aid  for Scientific Research (Nos. 
18$\cdot$2670, 16540249, 17540231, and 17540244) 
from the Japan Society for the Promotion of Science.
We thank the Yukawa Institute for Theoretical Physics at
Kyoto University, discussions during the YITP workshop YITP-W-05-01, on
``New Developments in Nuclear Self-Consistent Mean-Field Theories'' 
were useful in completing this work.
We also thank the Institute for Nuclear Theory at the University of Washington
for its hospitality and the Department of Energy for partial support during 
the completion of this work.


\appendix


\newpage



\newpage

\newpage

\newpage

\newpage

\newpage

\newpage


\newpage


\begin{thebibliography}{99}

\bibitem{row76} 
D. J. Rowe and R. Bassermann, Canad. J. Phys. {\bf 54} (1976), 1941.


\bibitem{bri76} 
D. M. Brink, M. J. Giannoni and M. Veneroni, 
Nucl. Phys. A {\bf 258} (1976), 237. 

\bibitem{vil77} 
F. Villars, Nucl. Phys. A {\bf 285} (1977), 269. 

\bibitem{mar77} 
T. Marumori, Prog. Theor. Phys. {\bf 57} (1977), 112. 

\bibitem{bar78} 
M. Baranger and M. Veneroni, Ann. of Phys. {\bf 114} (1978), 123.

\bibitem{goe78} 
K. Goeke and P.-G. Reinhard, Ann. of Phys. {\bf 112} (1978), 328.

\bibitem{mar80} 
T. Marumori, T. Maskawa, F. Sakata and  A. Kuriyama,
Prog. Theor. Phys. {\bf 64} (1980), 1294.

\bibitem{gia80} 
M. J. Giannoni and P. Quentin, Phys. Rev. C {\bf 21} (1980), 2060; 
Phys. Rev. C {\bf 21} (1980), 2076.

\bibitem{dob81} 
J. Dobaczewski and J. Skalski, Nucl. Phys. A {\bf 369} (1981), 123. 

\bibitem{goe81}
K. Goeke,  P.-G. Reinhard, and D. J. Rowe, Nucl. Phys. A {\bf 359}
	(1981), 408. 

\bibitem{muk81} 
A. K. Mukherjee and M. K. Pal, Phys. Lett. B {\bf 100} (1981), 457;
Nucl. Phys. A {\bf 373} (1982), 289.

\bibitem{row82} 
D. J. Rowe, Nucl. Phys. A {\bf 391} (1982), 307.

\bibitem{fio83} 
C. Fiolhais and R. M. Dreizler, Nucl. Phys. A {\bf 393} (1983), 205.


\bibitem{rei84}
P.-G. Reinhard, F. Gr\"{u}mmer and K. Goeke,  
Z. Phys. A {\bf 317} (1984), 339.  

\bibitem{kur83}
A. Kuriyama and M. Yamamura, 
Prog. Theor. Phys. {\bf 70} (1983), 1675; {\bf 71} (1984) 122.

\bibitem{yam84}
M. Yamamura, A. Kuriyama and S. Iida, Prog. Theor. Phys. {\bf 71}
	(1984), 109.

\bibitem{mat85}
M. Matsuo and K. Matsuyanagi, Prog. Theor. Phys. {\bf 74} (1985), 288.

\bibitem{mat86} 
M. Matsuo, Prog. Theor. Phys. {\bf 76} (1986), 372.

\bibitem{shi87} 
Y. R. Shimizu and K. Takada, Prog. Theor. Phys. {\bf 77} (1987), 1192.

\bibitem{yam87}
M. Yamamura and A. Kuriyama, Prog. Theor. Phys. Suppl. No. {\bf 93}
	(1987). 

\bibitem{bul89}
A. Bulgac, A. Klein, N. R. Walet and G. Do Dang, 
Phys. Rev. C {\bf 40} (1989), 945.

\bibitem{wal91}
N. R. Walet, G. Do Dang, and A. Klein, Phys. Rev. C {\bf 43} (1991),
	2254.

\bibitem{kle91a} 
A. Klein, N. R. Walet and G. Do Dang, Ann. of Phys. {\bf 208} (1991),
	90. 

\bibitem{kan94}
K. Kaneko, Phys. Rev. C {\bf 49} (1994), 3014.

\bibitem{nak98a}
T. Nakatsukasa and N. R. Walet, Phys. Rev. C {\bf 57} (1998), 1192.

\bibitem{nak98b}
T. Nakatsukasa and N. R. Walet, Phys. Rev. C {\bf 58} (1998), 3397. 

\bibitem{nak00}
T. Nakatsukasa, N. R. Walet and G. Do Dang, 
Phys. Rev. C {\bf 61} (1999), 014302.

\bibitem{lib99}
J. Libert, M. Girod and J.-P. Delaroche, Phys. Rev. C {\bf 60} (1999),
	054301.
 
\bibitem{yul99}
E. Kh. Yuldashbaeva, J. Libert, P. Quentin and M. Girod, 
Phys. Lett. B {\bf 461} (1999), 1.

\bibitem{pro04}
L. Pr\'ochniak, P. Quentin, D. Samsoen and J. Libert, 
Nucl. Phys. A {\bf 730} (2004), 59.

\bibitem{alm04a}
D. Almehed and N. R. Walet,
Phys. Rev. C {\bf 69} (2004), 024302.

\bibitem{alm04b}
D. Almehed and N. R. Walet,
Phys. Lett. B {\bf 604} (2004), 163.

\bibitem{kle91b}
A. Klein and E. R. Marshalek, Rev. Mod. Phys. {\bf 63} (1991), 375.

\bibitem{dan00}
G. Do Dang, A. Klein, and N. R. Walet, Phys. Rep. {\bf 335} (2000), 93.

\bibitem{kur01}
A. Kuriyama, K. Matsuyanagi, F. Sakata, K. Takada
and M. Yamamura (Ed.), 
Prog. Theor. Phys. Suppl. No. {\bf 141} (2001).

\bibitem{rin80}
P. Ring and P. Schuck, {\it The Nuclear Many-Body Problem} 
(Springer-Verlag, 1980).

\bibitem{bla86} 
J.-P. Blaizot and G. Ripka, {\it Quantum Theory of Finite Systems} 
(The MIT press, 1986).

\bibitem{abe83}
(Ed.) A. Abe and T. Suzuki, 
Prog. Theor. Phys. Suppl. Nos. {\bf 74} \&{\bf 75} (1983).

\bibitem{bri05}
D. M. Brink and R. A. Broglia, {\it Nuclear Superfluidity, 
Pairing in Finite Systems} 
(Cambridge University Press, 2005).
\bibitem{mat84}
M. Matsuo, Prog. Theor. Phys. {\bf 72} (1984) 666. 

\bibitem{mat85a}
M. Matsuo and K. Matsuyanagi, Prog. Theor. Phys. {\bf 74} (1985) 1227; 
Prog. Theor. Phys. {\bf 76} (1986), 93; 
Prog. Theor. Phys. {\bf 78} (1987), 591. 

\bibitem{mat85b} 
M. Matsuo, Y. R. Shimizu and K. Matsuyanagi,
  {\it Proceedings of The Niels Bohr Centennial Conf. on Nuclear
	Structure},
  ed. R.~Broglia, G.~Hagemann and B.~Herskind (North-Holland, 1985),
  p.~161.

\bibitem{tak89}
K. Takada, K. Yamada and H. Tsukuma, Nucl. Phys. A {\bf 496} (1989),
	224.

\bibitem{yam89a}
K. Yamada, K. Takada and H. Tsukuma, Nucl. Phys. A {\bf 496} (1989),
	239.

\bibitem{yam89b}
K. Yamada and K. Takada, Nucl. Phys. A {\bf 503} (1989), 53.

\bibitem{aib90}
H. Aiba,  Prog. Theor. Phys. {\bf 84} (1990), 908.
 
\bibitem{yam91}
K. Yamada, Prog. Theor. Phys. {\bf 85} (1991), 805; 
Prog. Theor. Phys. {\bf 89} (1993), 995.

\bibitem{ter91}
J. Terasaki, T. Marumori and F. Sakata, 
Prog. Theor. Phys. {\bf 85} (1991), 1235. 

\bibitem{ter92}
J. Terasaki, Prog. Theor. Phys. {\bf 88} (1992), 529;
Prog. Theor. Phys. {\bf 92} (1994), 535. 

\bibitem{mat92}
M. Matsuo, in {\it New Trends in Nuclear Collective Dynamics}, 
ed. Y. Abe, H. Horiuchi and K. Matsuyanagi,
(Springer-Verlag, 1992), p.219. 

\bibitem{shi01}
Y. R. Shimizu and K. Matsuyanagi, 
Prog. Theor. Phys. Suppl. No. {\bf 141} (2001), 285. 

\bibitem{mat00}
M. Matsuo, T. Nakatsukasa and K. Matsuyanagi, 
Prog. Theor. Phys. {\bf 103} (2000), 959.

\bibitem{kob04}
M. Kobayasi, T. Nakatsukasa, M. Matsuo and K. Matsuyanagi,
Prog. Theor. Phys. {\bf 112} (2004), 363;
Prog. Theor. Phys. {\bf 113} (2005), 129.

\bibitem{woo92}
J. L. Wood, K. Heyde, W. Nazarewicz, M. Huyse and P. van Duppen,
Phys. Rep. {\bf 215} (1992), 101.

\bibitem{fis00}
S. M. Fischer et al., Phys. Rev. Lett. {\bf 84} (2000), 4064;
Phys. Rev. C {\bf 67} (2003), 064318.

\bibitem{bou03}
E. Bouchez et al., Phys. Rev. Lett. {\bf 90} (2003), 082502.


\bibitem{hin06}
N. Hinohara, T. Nakatsukasa, M. Matsuo and K. Matsuyanagi,
Prog. Theor. Phys. 115 (2006) 567.

\bibitem{mat82}
K.~Matsuyanagi, Prog. Theor. Phys. {\bf 67} (1982), 1441;
{\it Proceedings of the Nuclear Physics Workshop}, Trieste, 5-30
	Oct. 1981.
ed. C. H. Dasso, R. A. Broglia and A. Winther (North-Holland, 1982),
p.~29. 

\bibitem{miz81}
Y.~Mizobuchi, Prog. Theor. Phys. {\bf 65} (1981), 1450.

\bibitem{suz88}
T.~Suzuki and Y.~Mizobuchi, Prog. Theor. Phys. {\bf 79} (1988), 480.

\bibitem{fuk91}
T.~Fukui, M.~Matsuo and K.~Matsuyanagi, Prog. Theor. Phys. {\bf 85}
	(1991), 281.
	
\bibitem{kob03}
M. Kobayasi, T. Nakatsukasa, M. Matsuo and K. Matsuyanagi,
Prog. Theor. Phys. {\bf 110} (2003), 65.
	
\bibitem{bar65}
M. Baranger and K. Kumar, Nucl. Phys. {\bf 62} (1965) 113; 
Nucl. Phys. A {\bf 110} (1968), 529; 
Nucl. Phys. A {\bf 122} (1968), 241;  
Nucl. Phys. A {\bf 122} (1968), 273. 

\bibitem{bar68}
M. Baranger and K. Kumar, Nucl. Phys. A {\bf 110} (1968), 490.

\bibitem{bes69} 
D. R. Bes and R. A. Sorensen,
{\it Advances in Nuclear Physics} vol. 2, (Prenum Press, 1969), p.~129.

\end{thebibliography}
\end{document}